\PassOptionsToPackage{pdfpagelabels=false}{hyperref}
\documentclass[fleqn,usenatbib]{mnras}
\usepackage{newtxtext,newtxmath}
\usepackage{amsmath}
\usepackage[flushleft]{threeparttable}
\usepackage{booktabs,caption}

\usepackage[T1]{fontenc}
\usepackage{ae,aecompl}
\usepackage{float}
\restylefloat*{figure}

\usepackage{xspace}
\usepackage{xcolor}
\usepackage{graphicx}	
\usepackage{multicol}
\usepackage{amsmath}	
\usepackage{ulem}
\usepackage{upgreek}
\usepackage{hyperref,xcolor}

\newcommand{\SO}{MAXI~J1820$+$070}

\title[Multiwavelength observations of MAXI~J1820$+$070]{Multiwavelength observations of MAXI~J1820$+$070 during its outburst decay and subsequent mini-outburst}

\author[M. \"{O}zbey Arabac{\i} et al.]{
M. \"{O}zbey Arabac{\i}$^{1}$\thanks{E-mail: m.ozbey-arabaci@soton.ac.uk},
E. Kalemci$^{2}$,
T. Din\c{c}er$^{3}$,
C. D. Bailyn$^{3}$,
D. Altamirano$^{1}$,
T. Ak$^{4}$,
\\
\\
$^{1}$Department of Physics and Astronomy, University of Southampton, Southampton, SO17 1BJ, UK\\
$^{2}$Faculty of Engineering and Natural Sciences, Sabanc{\i} University, Orhanl{\i}-Tuzla, 34956 Istanbul, Turkey\\
$^{3}$Department of Astronomy, Yale University, PO Box 208101, New Haven, CT 06520-8101, USA\\
$^{4}$Istanbul University, Faculty of Science, Department of Astronomy and Space Sciences, 34119, Istanbul, Turkey}

\date{Accepted XXX. Received YYY; in original form ZZZ}

\pubyear{2021}

\begin{document}
\label{firstpage}
\pagerange{\pageref{firstpage}--\pageref{lastpage}}
\maketitle

\begin{abstract}
We present results from quasi-simultaneous multiwavelength observations of the Galactic black hole X-ray transient MAXI J1820$+$070 during the decay of the 2018 outburst and its entire subsequent mini-outburst in March 2019. We fit the X-ray spectra with phenomenological and Comptonizaton models and discuss the X-ray spectral evolution comparing with the multiwavelength behaviour of the system. The system showed a rebrightening in UV/Optical/NIR bands 7-days after the soft-to-hard transition during the main outburst decay while it was fading in X-rays and radio. In contrast, the mini-outburst occurred 165-days after the hard state transition of the initial outburst decay and was detected in all wavelengths. For both events, the measured timescales are consistent with those observed in other black hole systems. Contemporaneous hard X-ray/soft $\gamma$-ray observations indicate a non-thermal electron energy distribution at the beginning of the UV/Optical/NIR rebrightening, whereas a thermal distribution can fit the data during the hard mini-outburst activity. The broadband spectral energy distributions until the rebrightening are consistent with the irradiated outer accretion disc model. However, both the SEDs produced for the peak of rebrightening and close to the peak of mini-outburst provided good fits only with an additional power-law component in the UV/Optical/NIR frequency ranges which is often interpreted with a jet origin.
\end{abstract}




\section{Introduction}
Galactic black hole transient (GBHT) systems spend most of their time in a faint, quiescent state where mass transfer rate from the accretion disc onto the black hole is at a very low level \citep{Tanaka95,McClintock06}. Occasionally, they become active and undergo weeks to months long transient outbursts \citep{Tanaka96, Dunn10, Reynolds13, Corral-Santana16, Tetarenko16} due to rapid and dramatic increase in mass accretion rate driven by the thermal-viscous instabilities developing in the disc \citep[][and references therein]{Meyer81,Coriat12}. During these outbursts, the GBHTs often follow a similar evolutionary track through a sequence of different X-ray states depending on their temporal and spectral properties. In general, a typical GBHT outburst starts in the hard state, make a transition to the soft state and eventually returns to the hard state at the end of the outburst \citep[see][for reviews]{Remillard06,Belloni10x,Belloni16}. In addition, during the transitions between the hard and soft states, the GBHT may go through several intermediate states displaying a cyclic ‘q-shaped’ pattern in the hardness-intensity diagram (HID) in which the X-ray luminosity is plotted versus spectral hardness \citep{Homan01,Maccarone03,Belloni05}.

In the hard state, the X-ray spectrum of a GBHT is dominated by a hard power-law component with a high energy cut-off broadly associated with the Comptonization of soft photons in a hot ($kT_{e}$ $\sim$ 100 keV), optically thin electron corona or hot inner flow \citep{Sunyaev79,Poutanen98, Gilfanov10,Zdziarski04}. The contribution from the optically thick regions to the X-ray emission in this state is very weak and generally interpreted as the truncation of the inner disc to be far away from the innermost stable circular orbit \citep[ISCO,][]{Esin97,Poutanen97}. By contrast, the inner disc radius may extend down or close to the ISCO in the soft state. The X-ray spectrum is characterized by a thermal, blackbody component peaking at $\sim$1 keV accompanied with a weak hard power-law tail extending well beyond the $\gamma$-ray regime without a break. The soft X-ray component is associated with a standard optically thick and geometrically thin accretion disc \citep{Shakura73} whereas the physical origin of the non-thermal high energy tail is still not clear \citep[see][for the application of different models for Cyg~X-1]{Cangemi21}. During the transitional intermediate states, the source spectrum is more complex compare to the main two states and includes both hard power-law emission from the corona/hot flow and soft thermal emission from the accretion disc \citep[see][and the references therein for the details of intermediate states and state transitions]{Remillard06}

 Multiwavelength monitoring of GBHTs have revealed that changes in the radio and optical-infrared (OIR) emission properties are closely related to the X-ray spectral states \citep{Fender09,Corbel00,Vadawale03,Homan05,Dincer12,MillerJones12,Russell12,RussellTD19,Kalemci13,Fender14,Carotenuto21}. In the hard state, the spectral energy distributions (SEDs) show a flat/slightly inverted synchrotron spectrum ($F_{\nu} \propto \nu^{\alpha}$, where the spectral index $\alpha$ $\geqslant$ 0) extending from radio to millimeter bands \citep{Fender_01,Tetarenko15} and breaking to an optically thin emission ($\alpha$ \textless ~0) at the IR regime indicating a collimated and compact jet \citep{Fender01,Russell06, Russell13, RussellTD14}. Close to the transition to the hard-to-soft state, however, the compact jet switches off and the radio emission is quenched below the detection levels in the soft state \citep{Fender99,Gallo03, RussellTD19, Carotenuto21}. Following the transition back to the hard state at the outburst decay, the compact jet reforms progressively as indicated by the evolution in the radio flux, radio spectral index and IR-optical SEDs \citep{MillerJones12, Kalemci13, Corbel13, RussellTD14}. Also in the OIR regime, correlations between X-ray and OIR flux have been determined \citep{Homan05, Coriat09}. This led to the discussion of possible X-ray emission mechanisms in the hard state other than thermal Comptonization, such as direct synchrotron \citep{Markoff01,Russell10} and/or synchrotron self-Compton radiation \citep{Markoff05} suggesting the jet could make significant contribution to the high frequency emission assuming that the hot electron corona as the base of the jet \citep{Markoff03}. A plausible way to probe this contribution on the X-ray spectra could be achieved by invoking the models including non-thermal/hybrid electron distribution in the spectral fits as the jet provides non-thermal electrons into the corona and modifies the electron energies.
 
The GBHT outburst light curves could be very complicated, and while the so called "the main outburst" could go through the spectral states described above, some GBHTs also show rebrightening episodes during the outburst decay \cite{Kalemci13} and/or an increase in brightness several days after the X-ray flux goes below the detection limits of the most observatories that are sometimes defined as mini-outbursts \citep{Chen97}. A systematic multiwavelength study of GBHTs in the outburst decay by \cite{Kalemci13} showed that for most of the systems, a rebrightening (secondary maximum or secondary flare) in OIR occurred $\sim$ 1-2 weeks after the soft-to-hard transition. Detection of rebrightening during the outburst decay supports the argument that formation of compact jet and its interaction with the accretion environment are imprinted on the multiwavelength behaviour of the GBHTs \citep{Buxton04,Buxton12,Kalemci05,Kalemci13,Dincer12, Corbel13}. Alternatively, the synchrotron radiation from the hot accretion flow model \citep{Poutanen98,Veledina13}, or the irradiation from the secondary star or outer part of the disc could explain the brightness increase in the OIR bands. In contrast, there are limited number of pointed hard X-ray observations for the mini-outbursts (e.g. XTE~J1752$-$223, SWIFT~J1745$-$26, and V404 Cyg,~\citealt{Chun13, Kalemci14, Munoz17}) since they have been observed frequently in the soft X-rays and optical (see \citealt{Chen97} for some historical examples, both in black holes and neutron stars). A recent study by \cite{Zhang19} attempted a classification of the rebrightenings during/after the main outburst decay based on the the available fluxes and applied this scheme to Swift J1753.5$-$0127 which showed a mini-outburst in radio, optical and X-rays. It can be seen that different flavors exist depending on whether the source reaches quiescence first. Some sources show multiple mini-outbursts after the initial outburst (e.g. XTE~J1650$-$500, MAXI~J1535$-$571~\citealt{Tomsick03, Cuneo20}). Although the origin of the mini-outbursts is still debated, an increased mass accretion triggered by the events during the evolution of the primary outburst through heating of the outer parts of the accretion disc \citep{Ertan02}, or the companion star \citep{Augusteijn93} are known to be the likely explanations. 

In this paper, we examine two such brightening episodes of \SO\ that occurred close to the end of its  main outburst decay in 2018 and the following post-outburst event in March 2019. Fig. \ref{fig:HID} shows the long-term hard X-ray (15-50 keV) light curve of \SO\ from the Burst Alert Telescope BAT onboard the \textit{Neil Gehrels Swift Observatory} \cite{Gehrels04} and corresponding HID (J. Wang, private communication, 2022) obtained from the \textit{Neutron Star Interior Composition Explorer} \cite[\textit{NICER}]{Gendreau16}. Previous works covering these episodes have used different classifications (e.g. rebrightening, brightening, flare, reflare or outburst) to signify the brightness increase in different wavelengths for MAXI~J1820$+$070. Here, we refer to the flux increase in the hard X-ray (before the state transition) and the UV-optical-near IR bands (UOIR) in the main outburst decay (during the hard state) as rebrightening and the one between MJD~58555-58590 as mini-outburst to be consistent with \cite{Kalemci13} and \cite{Zhang19}, respectively (see Fig.~\ref{fig:HID}). 

Determining the multiwavelength characteristics of GBHTs during and/or after the soft-to-hard transition at the final stages of the outburst is of great importance to disentangle the physical mechanisms governing these rebrightenings and understanding the conditions for the jet formation as well as the possible effects of jet on the X-ray spectral properties. Especially simultaneous and concurrent radio observations are extremely important as the origin of the optical emission can be originated from multiple sources, but the radio through infrared is generally established as being dominated by jet emission. Therefore, the work presented here mainly focuses on the hard state X-ray spectral properties and their evolution combined with multiwavelength information. Our data set includes quasi-simultaneous monitoring from \textit{Swift}, the Small \& Moderate Aperture Research Telescope  System \citep[SMARTS,][]{Subasavage10} and T\"{U}B\.{I}TAK National Observatory (TUG) at X-ray, ultraviolet (UV), optical, and near-infrared (NIR) bands, together with \textit{INTErnational Gamma-Ray Astrophysics Laboratory} \citep[\textit{INTEGRAL},][]{Winkler03} data in soft $\gamma$-rays, two covering the main outburst decay and the other one close to the peak of the mini-outburst which never reached the soft state. We also make use of the radio data collected from the previous reports \citep{Bright18,Bright20,Shaw21} to discuss the jet formation with respect to the evolution of the source as the radio emission is the best indicative of jet evolution. We should point out that this is the first multiwavelength study of MAXI~J1820$+$070 dedicated to discuss its multiwavelength behaviour extending to soft $\gamma$-ray regime (above 200 keV) at the main outburst decay as well as its subsequent mini-outburst. This paper is structured as follows: we introduce the source in Section~\ref{sect:source_bg} and  describe in detail our observations and data reduction procedures in Section~\ref{sect:observations}. Section~\ref{sect:results} presents the results of multiwavelength spectral evolution and broadband spectra. Finally, we discuss the implications of the results in Section~\ref{sect:discussion} and summarise them in Section~\ref{sect:sum}.

\section{MAXI~J1820+070}\label{sect:source_bg}
The X-ray transient \SO\ was discovered with the Monitor of All-Sky X-ray Image (MAXI) early in its outburst in March 2018 \citep{Kawamuro18} and it was soon associated with an optical transient ASASSN-18ey detected by the All-Sky Automated Survey for Super-Novae (ASAS-SN) 5 days prior to the X-rays \citep{Denisenko18}. Subsequent follow-up observations revealed that the optical/X-ray flux ratio and the X-ray spectral properties of the source to be consistent with those of GBHTs in the hard state \citep{Baglio18,Homan18,Uttley18}. This classification has been cemented dynamically by \cite{Torres19}. The system includes a BH with a mass of 5.95$^{0.39}_{0.22}$ M$_{\sun}$ and a K type companion \citep{Torres20} at a distance of 2.96$\pm$0.33 kpc \citep{Atri20}. 

A radio counterpart to the source was also detected with Arcminute Microkelvin Imager Large Array (AMI-LA) and the RATAN-600 telescope \citep{Bright18,Trushkin18} during the initial hard state that lasted more than 3 months \citep{Roques19,Shidatsu18}. Compact jet detection was reported by \cite{Bright18} and subsequent radio observations revealed the discovery of a short-lived radio flare associated with the launch of bi-polar superluminal ejections during the hard-to-soft state transition by \citep{Bright20}. \cite{Homan20} reported the switch between type-C to type-B quasi periodic oscillations (QPO) that might be linked to discrete ejections. These powerful ejection events were also detected in X-rays with \textit{Chandra} \citep{Espinasse20}. Using a new dynamic phase centre tracking technique on Very Long Baseline Array observations (VLBA) of the approaching fast moving ejecta, \cite{Wood21} identified a previously undetected approaching slow-moving jet knot lasted for $\sim$6 h. They showed that this slow component was responsible for the radio flare and for the QPO transition rather than the fast-moving ejecta stated in the previous works \citep{Bright20,Homan20}.   
 
\begin{figure*}
\hspace*{-0.55cm}
\centering
\includegraphics[scale=0.6]{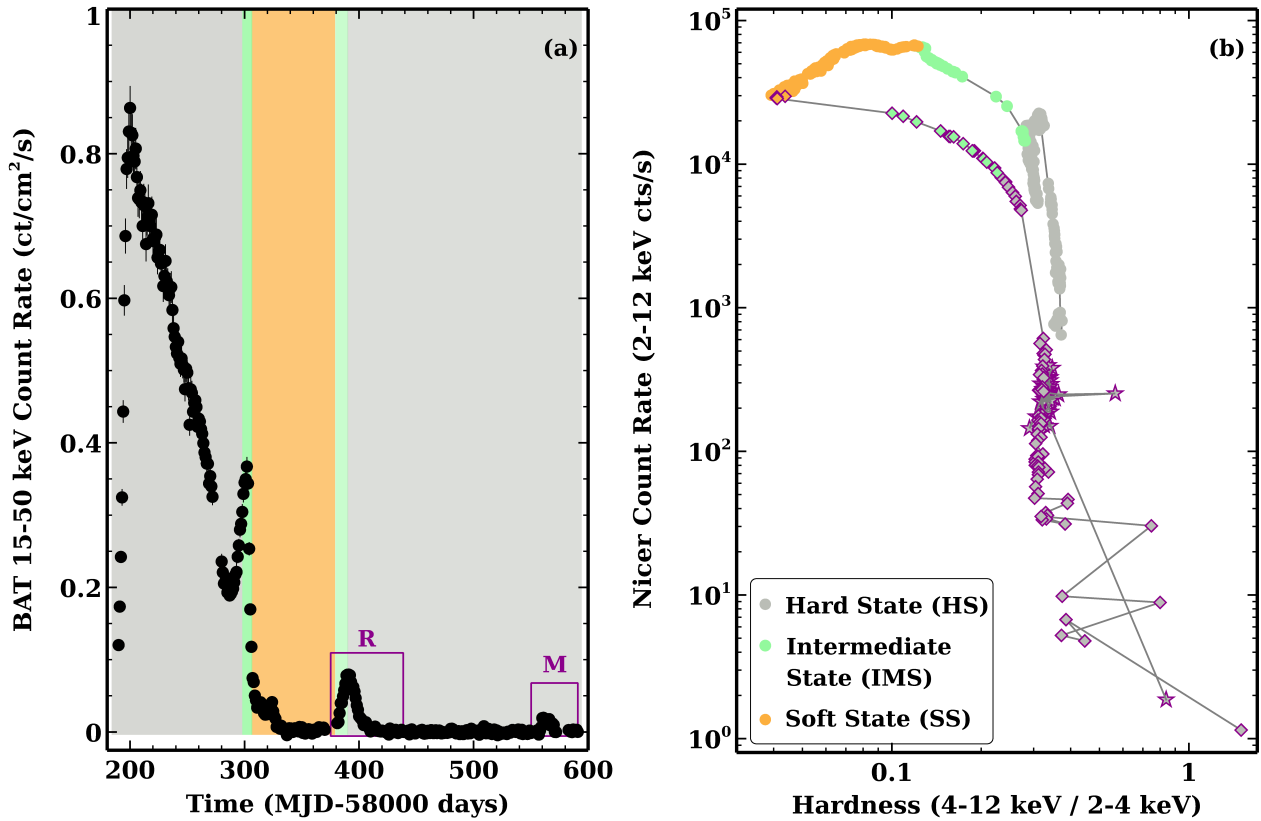}
\caption{(a) \textit{Swift}/BAT light curve of \SO\ covering the full outburst started in March 2018. The rebrightening (R) and mini-ouburst (M) epochs studied in this work are highlighted with purple boxes. Coloured areas represent the spectral states of the source and match with the HID produced from NICER data (J. Wang, private communication, 2022) in panel (b). The day coincident with the \textit{Swift}/XRT observations for the rebrightening and the mini-outburst events are denoted with the diamonds and stars in purple, respectively. }
\label{fig:HID}
\end{figure*}

The source went through all canonical accretion states during the 2018 outburst and underwent rebrightening episodes during the outburst decay before fading into quiescence in X-rays (or extremely low X-ray luminosities) in November 2018 \citep[][see also Fig.~\ref{fig:HID}]{Homan20,Shidatsu19} though pre-outburst level in the optical band was reached $\sim$2 months later \citep{Russell19}. After remaining in quiescence for $\sim$3.5 months, the source experienced a weak ($\sim$ ~30 mCrab in 2-4 keV band), 2-month long mini-outburst in March 2019 \citep{Ulowetz19,Bahramian19,Vozza19} and showed three more, relatively weak X-ray activities since then \citep{Xu19,Bright19,Hankins19,Adachi20,Sasaki20,Baglio21,Homan21,Sai21}. 

\begin{figure*}
\centering
\includegraphics[width=1.8\columnwidth]{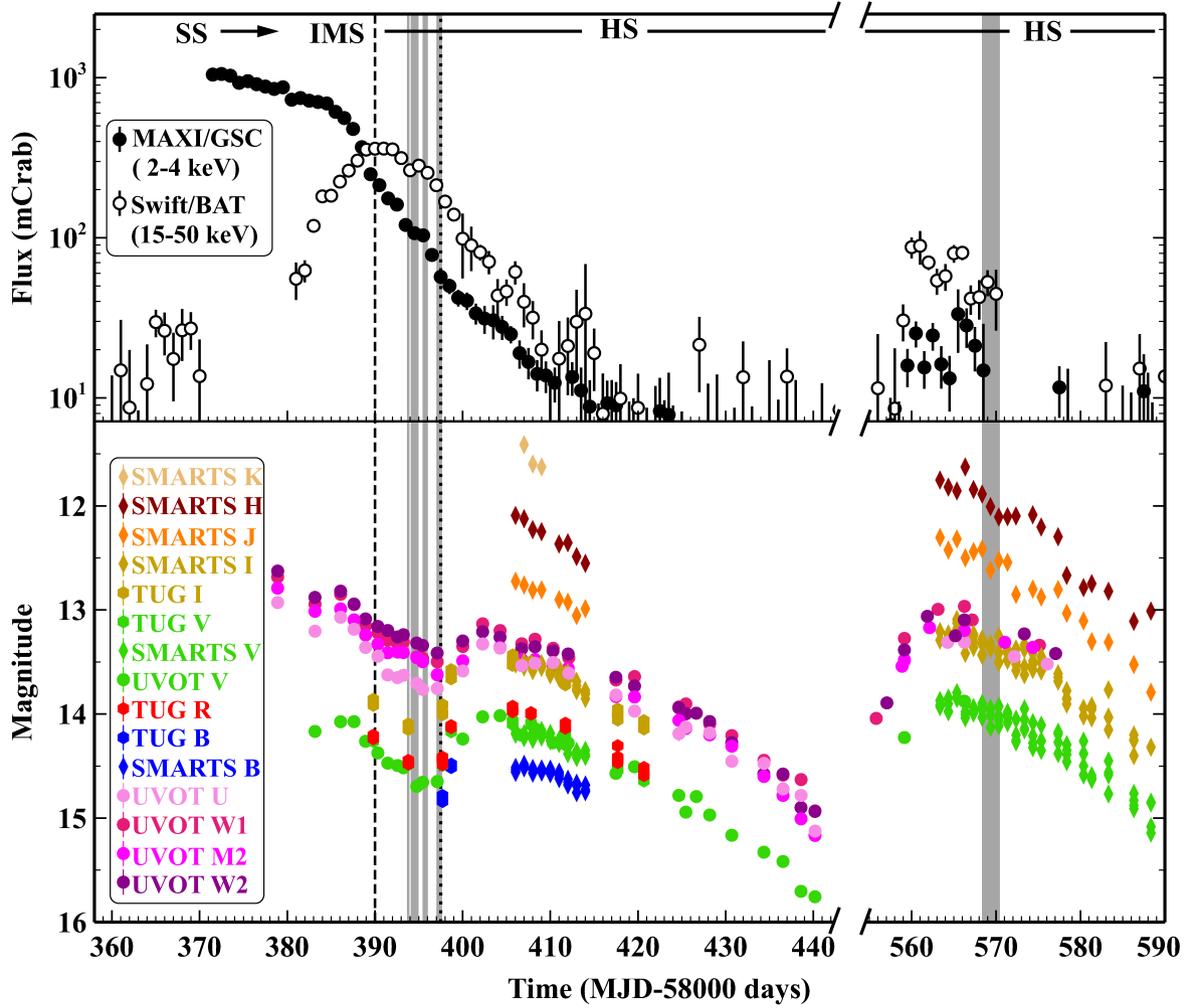}
\caption{Multiwavelength light curves of MAXI~J1820$+$070 during the rebrightening event in 2018 outburst decay (MJD~58360-58440) and the subsequent mini-outburst in 2019 (MJD~58555-58390). Dashed line separates the intermediate-to-hard state transition. Dotted line shows the onset of the rebrightening in UOIR band. Grey strips show the time of our \textit{INTEGRAL} observations given in Table~\ref{tab:INTEGRAL}. Note that space between Obs.1 and Obs.2 is added for the clarity. The break in the time axis corresponds to a gap of $\sim$110 days when the source was mainly in the quiescent state. }
\label{fig:uvoptir}
\end{figure*} 
\begin{figure*}
\begin{center}
\includegraphics[width=1.8\columnwidth]{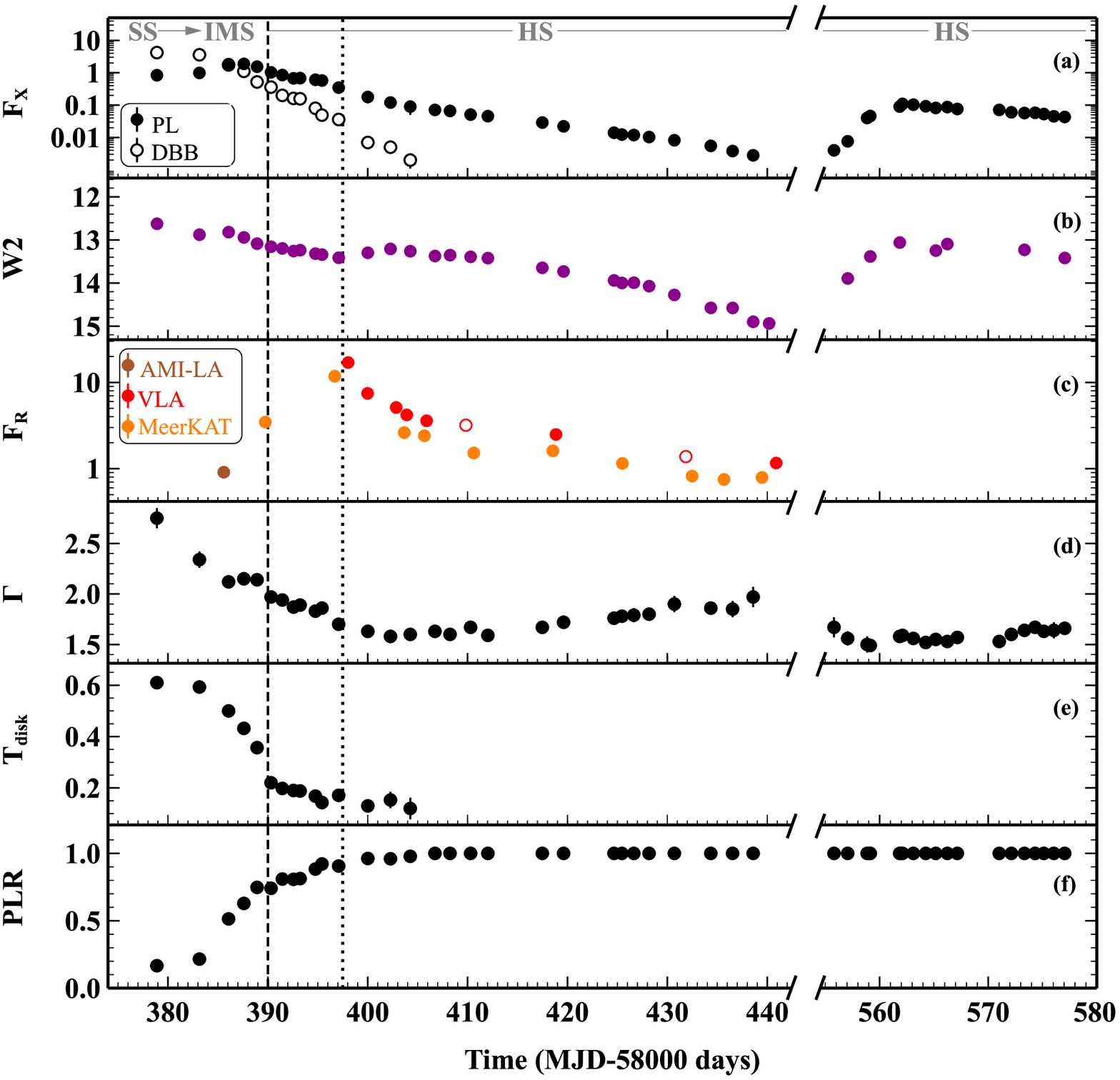}
\caption{Evolution of (a) the power-law and disc blackbody fluxes in the 0.5-10 keV energy range in units of 10$^{-8}$ erg s$^{-1}$ cm$^{-2}$, (b) the W2 magnitude, (c) flux densities at 1.28 GHz (MeerKAT), 6 GHz (VLA) and at 15.5 GHz (AMI-LA) in units of mJy, (d) the photon index of the power-law, (e) the disc inner temperature in keV, and (f) the ratio of the power-law flux to the total flux in the 0.5-10 keV band. The dashed and dotted lines match those in Fig.~\ref{fig:uvoptir}. For the radio observations; two VLA data points represented in open circles are taken from \protect\cite{Shaw21}, one AMI-LA detection obtained from \protect\cite{Bright18}, and the rest are from \protect\cite{Bright20}}.
\label{fig:multievol}
\end{center}
\end{figure*}
\section{Multiwavelength Observations and Analysis}\label{sect:observations}
Figure~\ref{fig:uvoptir} shows the 
soft X-ray (2-4 keV), hard X-ray (15-50 keV) and the UOIR band light curves of the outburst decay and the subsequent mini-outburst 
covered by MAXI/GSC \citep{Mihara11}, \textit{Swift}/BAT \citep{Barthelmy05}, SMARTS, and TUG telescopes. The soft and hard X-ray data are provided as daily averages by MAXI/GSC \citep{Matsuoka09} and \textit{Swift}/BAT teams \citep{Krimm13}. The dates of \textit{INTEGRAL} pointing observations are marked with grey vertical lines. As it is also shown in Fig.~\ref{fig:HID}, the source evolved from the soft to hard state during the rebrightening and remained in the hard state for the entire mini-outburst period.

\subsection{Swift}\label{sect:Swiftred}
We analysed a total of 47 \textit{Swift} observations collected with the XRT \citep{Burrows05} and UVOT \citep{Roming05} instruments between 17 September 2018 and 4 April 2019. The observations were taken exclusively during the outburst decay and mini-outburst phases, with a cadence of every few days.

The XRT spectra were extracted using the standard HEASOFT v6.24 \footnote{\url{https://www.swift.ac.uk/analysis/xrt/spectra.php}} with the 2018 July version of the HEASARC calibration database (CALDB). The first nine observations suffered from photon pile-up. To mitigate the pile-up problem, we limited our spectral extraction to the single pixel events and used an annulus region centred at the source location, excising the central pixels. More specifically, the inner radius of the annulus was set to the radius where the extracted count rate fell below 100 cts s$^{-1}$ and the outer radius of the annulus was fixed at 70\arcsec~(30 pixels). Spectra from the pile-up free observations were extracted in the same manner but without excluding the central pixels. For the instrumental calibration, the appropriate response matrix file \texttt{swxwt0s6psf1$\_$20131212v001.rmf} was obtained from HEASARC CALDB and auxiliary response files were created using \texttt{xrtmkarf} with the exposure map created by \texttt{xrtexpomap}. For model fitting, each spectrum was grouped to have at least one photon per spectral bin using \texttt{grppha}. 

UVOT photometry was done using the \texttt{uvotsource} tool, which returns both the background-corrected magnitudes in the Vega system and the flux densities in mJy. The source counts were extracted from a circular region with an aperture radius of 5$\arcsec$ centred on the source and the background counts were done from a source-free, circular region with a radius of 10$\arcsec$.
\begin{table*}
\begin{threeparttable}
\caption{Details of \textit{INTEGRAL} observations.}
\label{tab:INTEGRAL}
\begin{tabular}{lccccc}
\toprule
Obs. No & ObsID & Revolution & Start-End Date & ISGRI Exposure & \textit{Swift} ObsID \\
& & (MJD-58000) & (ks)&\\
(1) & (2) & (3) & (4) & (5)&(6)  \\
\midrule
1 & 15400050001 & 2006 & 393.62-394.01 &20.3 &00010627106\\
2 & 15400050001 & 2006 & 394.01-395.00 &51.6&00010627107\\
3 & 15400050001 & 2007 & 395.45-396.03 &28.2 &00010627108\\
4 & 15400050001 & 2007 & 397.01-397.65 & 33.4&00010627109\\
5 & 15400050001 & 2072 & 568.26-570.50 &103.92 &00010627149 \\
\bottomrule
\end{tabular}
\begin{tablenotes}
\small
\item Column (1): Observation number assigned for the \textit{INTEGRAL} and \textit{Swift} observations matched. Column (2): Observation ID. Column (3): \textit{INTEGRAL} revolution. Column (4): Start and end date of \textit{INTEGRAL} observations (MJD-58000 days). Column (5): Effective exposure time simultaneous with \textit{Swift} observations. Column (6): \textit{Swift} observation ID within the given observing time.
\end{tablenotes}
\end{threeparttable}
\end{table*}
\subsection{INTEGRAL}\label{sect:INTEGRAL}
We observed MAXI~J1820$+$070 with JEM-X \citep{Lund03}, IBIS Soft Gamma-Ray Imager  \citep[ISGRI,][]{Lebrun03}, and SPI \citep{Vedrenne03} instruments onboard the \textit{INTEGRAL} satellite during the revolutions of 2006, 2007, and 2072. The data were processed using the Off-line Scientific Analysis (\texttt{OSA}) software version 11.0 provided by the \textit{INTEGRAL} Science Data Centre \citep[ISDC;][]{Courvoisier03}. The standard extraction algorithm was used to obtain JEM-X and IBIS/ISGRI spectra for each revolution. JEM-X spectra were extracted for an energy range of 3--35 keV corresponding to 16 channels while IBIS/ISGRI spectra were produced to have 53 energy bins in 30--350 keV band. In order to have optimum simultaneity with the \textit{Swift} observations within the revolution, we split the extracted spectra into five segments (see Table~\ref{tab:INTEGRAL}). The average spectrum of each group has been obtained through \texttt{spe$\_$pick} tool to get a better signal to noise ratio. In this work we considered only JEM-X1 and IBIS/ISGRI instruments as JEM-X1 provided better statistics than JEM-X2 and SPI did not yield a source detection above 200 keV.

\subsection{SMARTS \& TUG}
As part of our SMARTS X-ray binary program, we observed MAXI J1820+070 on a near-daily basis in the second half of 2018 October as well as between 2019 March 21 and 2019 May 2. The observations were taken with the ANDICAM\footnote{\url{https://www.astronomy.ohio-state.edu/ANDICAM/detectors.html}, for further information about the instrument.} instrument \citep{Depoy03} on the SMARTS 1.3-m telescope at the Cerro Tololo Inter-American Observatory (CTIO) using the standard KPNO Johnson-Cousin optical \textit{BVI} filters and standard CIT/CTIO \textit{JHK} filters for the NIR. A nightly observing sequence consisted of several exposures in each of the optical bands and seven dithered exposures in each of the infrared bands. The exposure times were 30 s in V, 50 s in I and B, and 30 s in each of the dithered infrared images. We reduced all of the data in IRAF \citep{Tody86, Tody93} following the standard procedures described in \cite{Buxton12}. We performed point spread function photometry on all reduced images with the DAOPHOT4 suite of programs \citep{Stetson87}, and then converted the magnitudes to the Vega system with respect to four nearby field stars, with absolute calibration via the standard stars in the RU 149 field \citep{Landolt92} on clear nights and the Two Micron All-Sky Survey catalog \citep{Skrutskie06} in optical and NIR. We also observed the source in optical bands with the T100 telescope at TUG through our DDT and TOO programs. These observations were taken on nine nights between 2018 September 28 and October 29 UT with exposure times of 60 s in B and V, and 40 s in R and I\footnote{See Table~\ref{tab:tug} and Table~\ref{tab:smarts} for the effective wavelengths of the filters.}. We extracted the source magnitudes from T100 images in the same way we did from the SMARTS images.

\begin{table*}
\begin{threeparttable}
\caption{Best-fit \texttt{tbabs $\times$ (diskbb + power law)} parameters obtained from the \textit{Swift}/XRT spectral analysis, fixing N$_H$ at 1.2 $\times$ 10$^{21}$ atom cm$^{-2}$}
\label{tab:SwiftonlyNhfixed}
\begin{tabular}{ccccccc}
\toprule
Obs. Id & Start Time & T$_{disk}$ & $\Gamma$ & F$_{disk}$ & F$_{PL}$ & $W$-stat/dof\\
 & (MJD, UTC) & (keV) &  & (10$^{-8}$ cgs) & (10$^{-8}$ cgs) & \\
(1) & (2) & (3) & (4) & (5) & (6) & (7)\\
\midrule
00010627097 & 58378.89 & 0.610 $\pm$ 0.007 & 2.75 $\pm$ 0.10 & 4.21 $\pm$ 0.07 & 0.84 $\pm$ 0.08 & 1.00\\ 
00010627098 & 58383.15 & 0.593 $\pm$ 0.007 & 2.34 $\pm$ 0.08 & 3.60 $\pm$ 0.06 & 0.99 $\pm$ 0.07 & 1.02\\ 
00010627100 & 58386.07 & 0.500 $\pm$ 0.009 & 2.12 $\pm$ 0.04 & 1.75 $\pm$ 0.06 & 1.85 $\pm$ 0.07 & 1.00\\ 
00010627101 & 58387.60 & 0.432 $\pm$ 0.008 & 2.15 $\pm$ 0.04 & 1.11 $\pm$ 0.05 & 1.88 $\pm$ 0.06 & 0.99\\ 
00088657010 & 58388.92 & 0.357 $\pm$ 0.005 & 2.14 $\pm$ 0.02 & 0.52 $\pm$ 0.03 & 1.54 $\pm$ 0.03 & 1.24\\ 
00010627102 & 58390.32 & 0.220 $\pm$ 0.005 & 1.97 $\pm$ 0.03 & 0.36 $\pm$ 0.02 & 1.03 $\pm$ 0.02 & 1.07\\ 
00010627104 & 58391.45 & 0.198 $\pm$ 0.005 & 1.94 $\pm$ 0.02 & 0.20 $\pm$ 0.01 & 0.85 $\pm$ 0.01 & 1.07\\ 
00010627105 & 58392.57 & 0.190 $\pm$ 0.005 & 1.87 $\pm$ 0.02 & 0.161 $\pm$ 0.008 & 0.675 $\pm$ 0.007 & 1.17\\ 
00010627106 & 58393.23 & 0.188 $\pm$ 0.006 & 1.89 $\pm$ 0.03 & 0.157 $\pm$ 0.008 & 0.683 $\pm$ 0.008 & 1.06\\ 
00010627107 & 58394.76 & 0.168 $\pm$ 0.011 & 1.83 $\pm$ 0.03 & 0.080 $\pm$ 0.008 & 0.607 $\pm$ 0.008 & 1.06\\
00010627108 & 58395.42 & 0.143 $\pm$ 0.009 & 1.86 $\pm$ 0.02 & 0.049 $\pm$ 0.004 & 0.575 $\pm$ 0.004 & 1.20\\ 
00010627109 & 58397.09 & 0.171 $\pm$ 0.021 & 1.70 $\pm$ 0.02 & 0.036 $\pm$ 0.003 & 0.348 $\pm$ 0.003 & 0.95\\ 
00010627110 & 58400.01 & 0.130 $\pm$ 0.022 & 1.63 $\pm$ 0.03 & 0.007 $\pm$ 0.002 & 0.178 $\pm$ 0.003 & 0.95\\ 
00010627111 & 58402.28 & 0.153 $\pm$ 0.032 & 1.58 $\pm$ 0.04 & 0.005 $\pm$ 0.001 & 0.120 $\pm$ 0.002 & 1.08\\ 
00010627112 & 58404.26 & 0.120 $\pm$ 0.042 & 1.60 $\pm$ 0.03 & 0.0020 $\pm$ 0.0009 & 0.09 $\pm$ 0.04 & 0.98\\ 
00010627113 & 58406.73 & - & 1.63 $\pm$ 0.02 & - & 0.071 $\pm$ 0.001 & 0.98\\ 
00010627114 & 58408.24 & - & 1.60 $\pm$ 0.02 & - & 0.066 $\pm$ 0.001 & 0.92\\ 
00010627115 & 58410.30 & - & 1.67 $\pm$ 0.03 & - & 0.051 $\pm$ 0.001 & 1.10\\ 
00010627116 & 58412.03 & - & 1.59 $\pm$ 0.02 & - & 0.0455 $\pm$ 0.0009 & 0.97\\ 
00010627119 & 58417.48 & - & 1.67 $\pm$ 0.02 & - & 0.0289 $\pm$ 0.0004 & 0.91\\ 
00010627120 & 58419.61 & - & 1.72 $\pm$ 0.02 & - & 0.0221 $\pm$ 0.0003 & 0.95\\ 
00010627122 & 58424.66 & - & 1.76 $\pm$ 0.04 & - & 0.0140 $\pm$ 0.0005 & 0.92\\ 
00010627123 & 58425.45 & - & 1.78 $\pm$ 0.05 & - & 0.0123 $\pm$ 0.0005 & 0.87\\ 
00010627124 & 58426.64 & - & 1.79 $\pm$ 0.07 & - & 0.0119 $\pm$ 0.0006 & 0.85\\ 
00010627125 & 58428.17 & - & 1.80 $\pm$ 0.04 & - & 0.0103 $\pm$ 0.0003 & 0.95\\ 
00010627126 & 58430.70 & - & 1.90 $\pm$ 0.08 & - & 0.0082 $\pm$ 0.0004 & 0.97\\ 
00010627128 & 58434.35 & - & 1.86 $\pm$ 0.06 & - & 0.0055 $\pm$ 0.0003 & 0.90\\ 
00010627129 & 58436.53 & - & 1.85 $\pm$ 0.08 & - & 0.0038 $\pm$ 0.0002 & 0.94\\ 
00010627130 & 58438.59 & - & 1.97 $\pm$ 0.10 & - & 0.0028 $\pm$ 0.0002 & 0.93\\ 
00010627136 & 58555.80 & - & 1.67 $\pm$ 0.10 & - & 0.0040 $\pm$ 0.0003 & 0.86\\ 
00010627139 & 58557.06 & - & 1.56 $\pm$ 0.07 & - & 0.0076 $\pm$ 0.0005 & 0.89\\ 
00010627140 & 58558.85 & - & 1.50 $\pm$ 0.08 & - & 0.040 $\pm$ 0.003 & 0.94\\ 
00010627141 & 58559.14 & - & 1.49 $\pm$ 0.05 & - & 0.046 $\pm$ 0.002 & 0.93\\ 
00010627143	& 58561.85 & - & 1.58 $\pm$ 0.02 & - & 0.090 $\pm$ 0.002 & 1.03\\ 
00010627144 & 58562.11 & - & 1.59 $\pm$ 0.02 & - & 0.109 $\pm$ 0.002 & 0.98\\ 
00010627145 & 58563.11 & - & 1.56 $\pm$ 0.02 & - & 0.103 $\pm$ 0.002 & 1.07\\ 
00010627146 & 58564.25 & - & 1.52 $\pm$ 0.02 & - & 0.092 $\pm$ 0.002 & 0.98\\ 
00010627147 & 58565.17 & - & 1.55 $\pm$ 0.02 & - & 0.082 $\pm$ 0.002 & 0.93\\ 
00010627148 & 58566.24 & - & 1.53 $\pm$ 0.02 & - & 0.087 $\pm$ 0.001 & 1.04\\ 
00010627149 & 58567.16 & - & 1.57 $\pm$ 0.02 & - & 0.076 $\pm$ 0.001 & 1.00\\ 
00010627150 & 58571.02 & - & 1.53 $\pm$ 0.03 & - & 0.071 $\pm$ 0.002 & 0.95\\ 
00010627151	& 58572.14 & - & 1.60 $\pm$ 0.03 & - & 0.060 $\pm$ 0.001 & 0.99\\ 
00010627152 & 58573.34 & - & 1.64 $\pm$ 0.03 & - & 0.057 $\pm$ 0.001 & 1.01\\ 
00010627153 & 58574.32 & - & 1.67 $\pm$ 0.03 & - & 0.058 $\pm$ 0.001 & 1.08\\ 
00010627154 & 58575.13 & - & 1.63 $\pm$ 0.03 & - & 0.053 $\pm$ 0.001 & 0.90\\ 
00010627155 & 58576.05 & - & 1.64 $\pm$ 0.08 & - & 0.045 $\pm$ 0.003 & 0.90\\ 
00010627156	& 58577.05 & - & 1.66 $\pm$ 0.04 & - & 0.043 $\pm$ 0.001 & 0.81\\ 
\bottomrule
\end{tabular}
\begin{tablenotes}
\small
\item Column (1): Observation Id. Column (2): Observation start time (MJD = JD-2400000.5). Column (3): disc inner temperature. Column (4): Photon index of the power-law. Column (5): disc flux in the 0.5-10 keV band. (6): Power-law flux in the 0.5-10 keV band. Column (7): Ratio of W-statistic to degrees of freedom. Errors on the fit parameters refer to the 1$\sigma$ uncertainties.
\end{tablenotes}
\end{threeparttable}
\end{table*}

\begin{table*}
\begin{threeparttable}
\caption{Best-fit \texttt{highecut $\times$ power law} parameters obtained from the XRT + JEM-X + ISGRI spectral analysis}
\label{tab:highepow}
\begin{tabular}{ccccccccc}
\toprule
Obs No & C$_{JEM-X}$ & C$_{ISGRI}$ & T$_{disk}$ & $\Gamma$ & E$_{fold}$ & E$_{cut}$ & Flux & $\chi^2 (\nu)$ \\
 & & & (keV) & & (keV) & (keV) & (10$^{-8}$ cgs) &  \\
(1) & (2) & (3) & (4) & (5) & (6) & (7) & (8) & (9)\\
\midrule
1 & 0.78 $\pm$ 0.02 & 1.05 $\pm$ 0.06 & 0.200 $\pm$ 0.005 & 1.79 $\pm$ 0.02 & 154$^{+56}_{-39}$ & 99.79 $\pm$ 9.60 & 2.18 $\pm$ 0.02 & 1.23 (552) \\
2 & 0.81 $\pm$ 0.02 & 0.99 $\pm$ 0.06 & 0.197 $\pm$ 0.009 & 1.74 $\pm$ 0.02 & 199$^{+34}_{-29}$ & 84.13 $\pm$ 6.33 & 1.99 $\pm$ 0.02 & 1.32 (445) \\
3 & 0.75 $\pm$ 0.02 & 1.15 $\pm$ 0.06 & 0.167 $\pm$ 0.008 & 1.80 $\pm$ 0.02 & 136$^{+64}_{-50}$ & 117.32 $\pm$ 12.81 & 1.83 $\pm$ 0.02 & 1.53 (589) \\
4 & 0.92 $\pm$ 0.03 & 1.13 $\pm$ 0.06 & 0.175 $\pm$ 0.008 & 1.70 $\pm$ 0.02 & 219$^{+63}_{-46}$ & 83.13 $\pm$ 8.50 & 1.37 $\pm$ 0.01 & 1.00 (571) \\
5 & 0.84 $\pm$ 0.04 & 0.83 $\pm$ 0.06 & - & 1.60 $\pm$ 0.02 & 202$^{+182}_{-81}$ & $<$103\tnote{$\dagger$} & 0.29 $\pm$ 0.01 & 1.08 (327)\\
\bottomrule
\end{tabular}
\begin{tablenotes}
\small
\item Column (1): Observation number. Column (2-3): Energy independent cross-instrument normalization factors for the JEM-X and ISGRI, . C$_{XRT}$ was frozen at 1. Column (4): disc inner temperature in units of keV. Column (5): Photon index ($\Gamma$) of the power-law. Column (6): Folding energy in the highecut model. Column (7): Cut off energy in the highecut model. Column (8): 0.5-350 keV unabsorbed flux in units of 10$^{-8}$ erg cm$^{-2}$ s$^{-1}$. Column (9): Reduced $\chi^2$ for $\nu$ degree of freedom.
\item[$\dagger$]$ 3\sigma$ upper limit.
\end{tablenotes}
\end{threeparttable}
\end{table*}

\begin{table*}
\begin{threeparttable}
\caption{Best-fit \texttt{const $\times$ tbabs (diskbb + compps)} parameters obtained from the XRT + JEM-X + ISGRI spectral analysis}
\label{tab:diskbbcompps}
\begin{tabular}{cccccccc}
\toprule
Obs No & C$_{JEM-X}$ & C$_{ISGRI}$ & T$_{disk}$/T$_{seed}$ & $\tau$ & kT$_e$ & $\chi^2 (\nu)$ \\
 & & & (keV) & & (keV) & &\\
(1) & (2) & (3) & (4) & (5) & (6) & (7) \\
\midrule
1 & 0.75 $\pm$ 0.05 & 0.92 $\pm$ 0.01 & 0.204 $\pm$ 0.004 & 2.20$^{+0.20}_{-0.30}$ & 67 $\pm$ 8 & 1.33 (553) \\
2 & 0.78 $\pm$ 0.02 & 0.84 $\pm$ 0.04 & 0.202 $\pm$ 0.008 & 2.87$^{+0.13}_{-0.25}$ & 57 $\pm$ 4 & 1.44 (447) \\
3 & 0.72 $\pm$ 0.02 & 0.91 $\pm$ 0.05 & 0.187 $\pm$ 0.008 & 1.88$^{+0.30}_{-0.17}$ & 78 $\pm$ 8 & 1.77 (589) \\
4 & 0.90 $\pm$ 0.03 & 1.06 $\pm$ 0.06 & 0.177 $\pm$ 0.007 & 2.42$^{+0.43}_{-0.37}$ & 67 $\pm$ 10 & 1.03 (572) \\
5 & 0.73 $\pm$ 0.03 & 0.80 $\pm$ 0.06 & 0.100 & 3.00$_{-0.37}$ & 65 $\pm$ 5 & 1.10 (331)  \\
\bottomrule
\end{tabular}
\begin{tablenotes}
\small
\item Column (1): Observation number. Column (2-3): Energy independent cross-instrument normalization factors for the JEM-X and ISGRI, . C$_{XRT}$ was frozen at 1. Column (4): disc inner temperature in units of keV. Column (5): Optical depth. Column (6): Electron temperature in units of keV. Column (7): Reduced $\chi^2$ for $\nu$ degree of freedom.
\end{tablenotes}
\end{threeparttable}
\end{table*}

\begin{table*}
\begin{threeparttable}
\caption{Best-fit \texttt{const $\times$ tbabs (diskbb + eqpair)} parameters obtained from the XRT + JEM-X + ISGRI spectral analysis}
\label{tab:diskbbeqpair}
\begin{tabular}{cccccccc}
\toprule
Obs No & C$_{JEM-X}$ & C$_{ISGRI}$ & T$_{disk}$/T$_{seed}$ & $l_h/l_s$ & $l_{nth}/l_h$ & $\tau_p$ & $\chi^2 (\nu)$ \\
 & & & (keV) & & & &\\
(1) & (2) & (3) & (4) & (5) & (6) & (7) & (8) \\
\midrule
\multicolumn{8}{|c|}{Thermal Comptonization}\\
\midrule
1 & 0.74 $\pm$ 0.02 & 0.94 $\pm$ 0.04 & 0.202 $\pm$ 0.004 & 7.7$^{+0.1}_{-0.4}$ & 0 & 1.46$^{+0.07}_{-0.07}$ & 1.30 (553) \\
2 & 0.78 $\pm$ 0.02 & 0.89 $\pm$ 0.04 & 0.198 $\pm$ 0.007 & 9.3$^{+0.3}_{-0.3}$ & 0 & 1.67$^{+0.04}_{-0.04}$ & 1.42 (447) \\
3 & 0.71 $\pm$ 0.02 & 1.00 $\pm$ 0.04 & 0.178 $\pm$ 0.005 & 7.8$^{+0.4}_{-0.3}$ & 0 & 1.14$^{+0.05}_{-0.04}$ & 1.69 (589) \\
4 & 0.89 $\pm$ 0.03 & 1.05 $\pm$ 0.05 & 0.175 $\pm$ 0.010 & 10.6$^{+0.6}_{-0.7}$ & 0  & 1.68$^{+0.10}_{-0.13}$ & 1.02 (572) \\
5 & 0.72 $\pm$ 0.03 & 0.76 $\pm$ 0.07 & 0.100 & 20.1$^{+1.9}_{-0.8}$ & 0 & 2.54$^{+0.34}_{-0.19}$ & 1.09 (331)\\
\midrule
\multicolumn{8}{|c|}{Hybrid Comptonization}\\
\midrule
1  & 0.78 $\pm$ 0.02 & 0.99 $\pm$ 0.05 & 0.190 $\pm$ 0.005 & 8.4$^{+0.9}_{-0.5}$ & 1.00$_{-0.09}$ & 1.53$^{+0.14}_{-0.12}$ & 1.21 (552) \\
2 & 0.81 $\pm$ 0.02 & 0.88 $\pm$ 0.03 & 0.185 $\pm$ 0.008 & 10.8$^{+0.7}_{-0.3}$ & 1.00$_{-0.26}$ & 1.91$^{+0.13}_{-0.02}$ & 1.34 (446) \\
3  & 0.75 $\pm$ 0.02 & 1.04 $\pm$ 0.04 & 0.150 $\pm$ 0.004 & 8.5$^{+0.5}_{-0.2}$ & 1.00$_{-0.02}$ & 1.18$^{+0.07}_{-0.06}$ & 1.48 (588) \\
4  & 0.92 $\pm$ 0.03 & 1.02 $\pm$ 0.05 & 0.164 $\pm$ 0.008 & 12.4$^{+1.0}_{-0.9}$ & 1.00$_{-0.40}$ & 1.85$^{+0.20}_{-0.17}$ & 1.00 (571) \\
5 & 0.72 $\pm$ 0.03 & 0.75 $\pm$ 0.07 & 0.100 & 20.9$^{+1.3}_{-1.1}$ & 0.12 $\pm$ 0.03  & 2.50$^{+0.50}_{-0.35}$ & 1.09 (330) \\
\bottomrule
\end{tabular}
\begin{tablenotes}
\small
\item Column (1): Observation number. Column (2-3): Energy independent cross-instrument normalization factors for the JEM-X, and ISGRI, . C$_{XRT}$ was frozen at 1. Column (4): disc inner temperature in units of keV. Column (5): Ratio of hard to soft compactnesses. Column (6): Ratio of the power supplied to energetic particles which goes into accelerating non-thermal particles. Column (7): Thomson scattering depth. Column (8): Reduced $\chi^2$ for $\nu$ degree of freedom.
\end{tablenotes}
\end{threeparttable}
\end{table*}

\begin{figure*}
 \begin{multicols}{2}
  \includegraphics[width=\columnwidth]{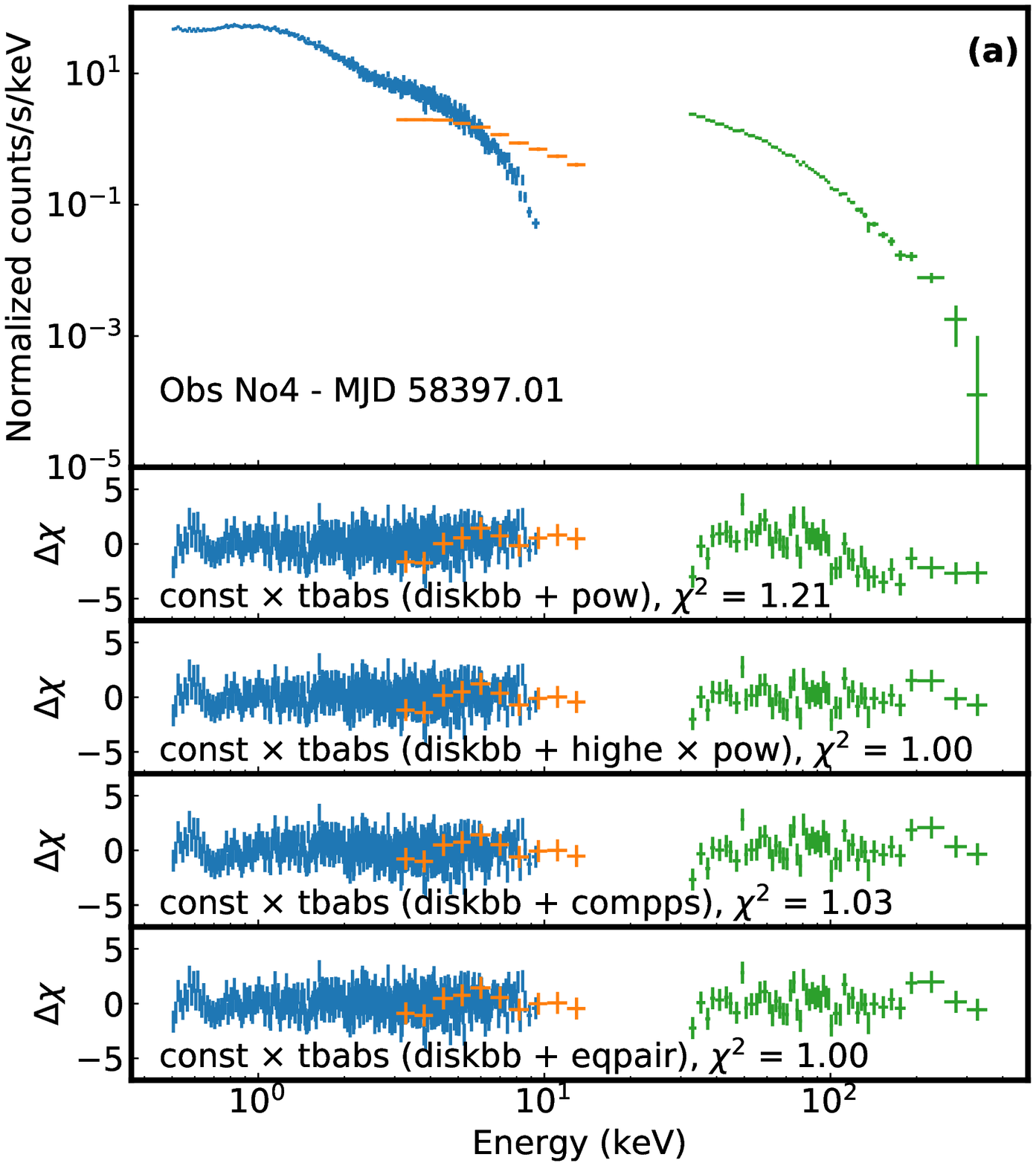}\par
  \includegraphics[width=\columnwidth]{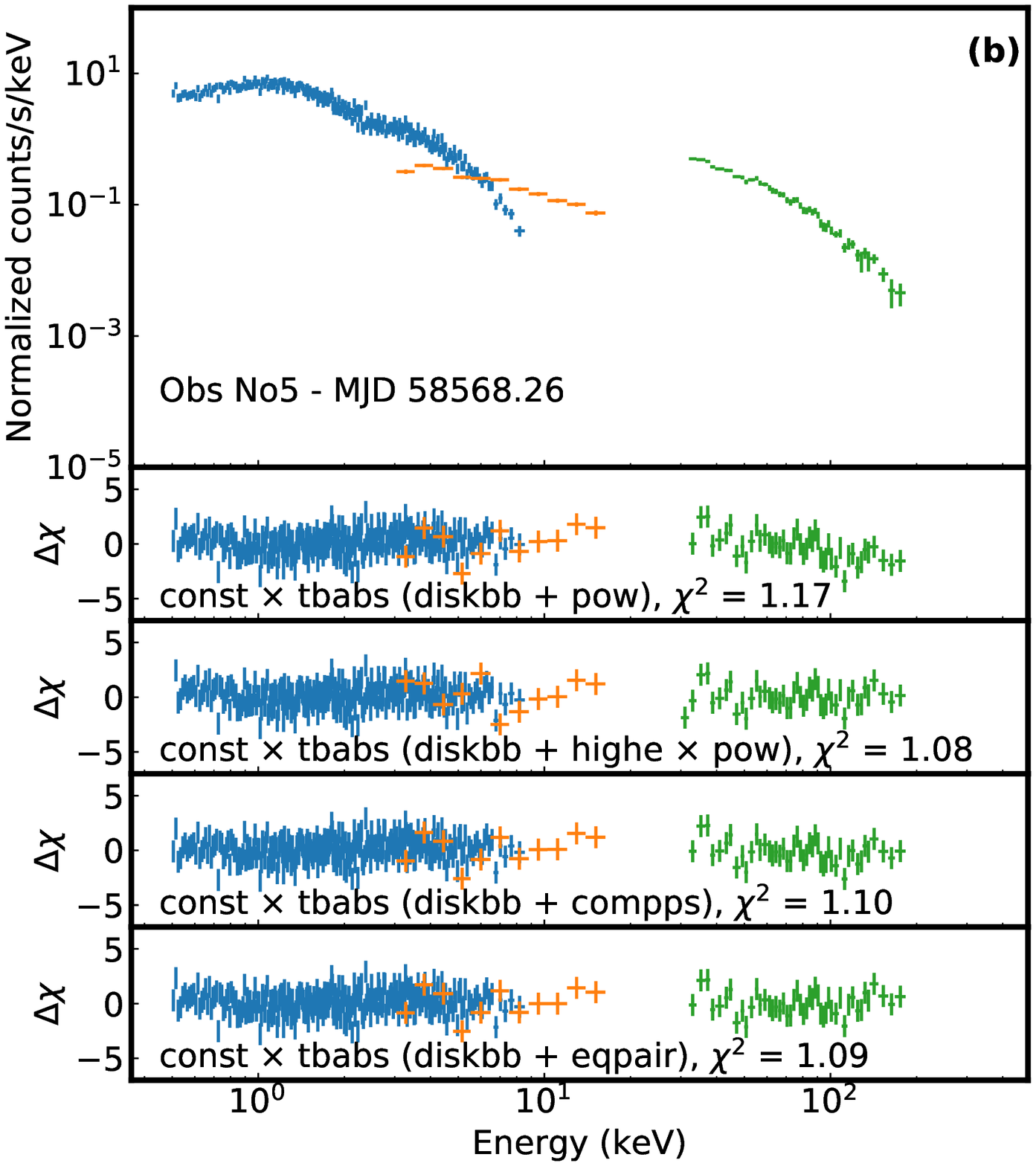}\par
  \end{multicols}
 \caption{Two broadband spectra compiled from quasi-simultaneous \textit{Swift}/XRT (blue), \textit{INTEGRAL} JEM-X (orange) and ISGRI (green) observations, along with $\Delta\chi$ = (Model - Data) / Error values for various models: (a) Onset of UOIR rebrightening, (b) mini-outburst.}
 \label{fig:broadband}
\end{figure*} 
 
 \section{Results} \label{sect:results}

\subsection{Multiwavelength Light Curves}
Figure~\ref{fig:uvoptir} presents the multiwavelength light curves of the main outburst decay and the mini-ouburst. The morphology of the hard X-ray and UOIR light curves are different for these two epochs. The rebrightening shows itself first in the hard X-rays between MJD~58380-58390. Afterwards, the UOIR band rebrightening occurs lasting until MJD~58440 while the hard X-ray flux drops. The mini-outburst, on the other hand, exhibits an increase in all bands after MJD~58555.

Determining the onset date of the rebrightening in the UOIR light curve is important as it allows investigation of associated changes in X-ray spectral and timing characteristics of the source around the same time for possible connections \citep{Kalemci13, Dincer14}. We were able to constrain the onset of the rebrightening with a higher precision using the method described in Appendix A.3. of \cite{Kalemci13}. Thanks to the cadence of UVOT observations, we had a superior coverage between MJD 58390 and 58403 compared to the previous reports \citep{Baglio18_atel} and we were able to fit before and during transition points with much smaller errors whereas \citealt{Shidatsu19} relies on $g'$ data with a large gap of points before the transition. We found that the rebrightening in V-band started on MJD 58397.5 $\pm$ 0.5 day, $~$a week after the hard state transition as shown in the next section.

We also note a possible rebrightening in the 10-20 keV band with MAXI between MJD~58402 and MJD~58410, almost coinciding with the UOIR peak. However, we cannot validate the statistical significance of this increase (adding a Gaussian peak to an exponentially decaying light curve provided a 0.10 chance probability with F-test). We only note that such increases in the hard X-rays are observed in GBHTs for those sources followed with pointed observations \citep{Kalemci13}. 

Finally it is worth comparing the evolution in radio to the evolution in UOIR (see Figs.~\ref{fig:uvoptir} and \ref{fig:multievol}). The radio flux increases after the initial measurement on MJD 58385.6 with AMI-LA \citep{Bright18} until MJD~58398, after which it starts to decrease with the X-ray flux. There is a delay of $\sim$ 7-day between the compact radio peak and the UOIR band peak. No radio data has been reported covering the mini-outburst.

\subsection{Spectral Results}\label{sect:xray}
All spectral analysis in the following subsections was performed using XSPEC v12.10.0c \citep{Arnaud96}. We took into account the Galactic absorption using the Tuebingen-Boulder absorption model \citep[a.k.a. \texttt{tbabs},][]{Wilms00} with the abundances of ~\cite{Wilms00} and the cross sections of ~\cite{Verner96}. We also used a systematic errors of 3 and 1 per cent in the fits for data from the JEM-X and ISGRI instruments, . All quoted uncertainties are at the 90\% confidence level unless otherwise stated.

\subsubsection{Multiwavelength Evolution: Spectral States}\label{sect:states}
We performed spectral fitting on the XRT data first to inspect the X-ray states of the source during its main outburst decay and the mini-outburst. To do this, we used a composite model consisting of a standard disc blackbody \citep[\texttt{diskbb},][]{Mitsuda84} and a \texttt{power law}. The model described most of the spectra very well but over-fit the faint ones taken with short exposures, and in return yielded largely unconstrained parameters if the $N_{H}$ values were kept free. In order to constrain the parameters better at all times, we first considered only the first four observations in our data set as they had the highest signal to noise ratio. The best-fit $N_H$ values from these observations were statistically consistent with each other around a mean of (1.2 $\pm$ 0.3) $\times$ 10$^{21}$ cm$^{-2}$ or a color excess of E(B-V) = 0.16 $\pm$ 0.05 \citep[via Eq. 15 in][]{Zhu17}. Having additional support from the fact that this color excess is consistent with the color excess of E(B-V) = 0.218 $\pm$ 0.003 reported in \cite{Schlafly11}, we fixed the N$_H$ at our average value and performed the fits again. The resulting best-fit parameters are shown in Table~\ref{tab:SwiftonlyNhfixed}. 

Figure~\ref{fig:multievol} presents the evolution of the X-ray spectral properties together with the U-band and the radio light curves of \SO. During the rise of the rebrightening in hard X-rays, between MJD 58378 and 58390, the spectra are generally soft with a mean photon index of 2.3 but a prominent change between the disc and the power-law components is also seen. In particular, the disc temperature decreases and the power-law flux peaks on MJD 58387.6. These changes indicate that the source was in transition from the soft-to-intermediate state. Note that we are not able to distinguish sub-intermediate states as we do not have fast X-ray timing information. However, in the following days, the source became detectable in the radio with a flux density of 3.47 mJy at 1.28 GHz \citep{Bright20} followed by a sharp drop both in the spectral index and the disc temperature on MJD 58390. We mark this date as the transition from the intermediate-to-hard state, after which the power-law component starts to dominate the spectrum substantially. After the transition, the power-law ratio (PLR, the ratio of the power-law flux to the total flux in the 0.5$-$10 keV band) increased to a value of 0.91 within seven days and a rebrightening occurs in all UOIR band on $\sim$ MJD 58397.5. As seen from the changes in the photon index, the spectrum gradually hardens and then starts softening after MJD 58402 while the X-ray flux continues decreasing. 
During the mini-outburst (starting at MJD~58555) UV and X-ray fluxes were correlated and the system remained in the hard state as indicated by the X-ray spectral power indices. 

\subsubsection{X-ray/$\gamma$-ray Spectra}
\label{sect:spectra}
As noted in Section~\ref{sect:INTEGRAL}, we have obtained five X-ray/$\gamma$-ray broadband spectra by combining simultaneous \textit{Swift} and \textit{INTEGRAL} observations. The first four spectra were taken in the hard state of the main outburst decay, corresponding to the before/onset of the UOIR rebrightening and the last one on top of the hard mini-outburst (Figure~\ref{fig:uvoptir}). We performed fits in the soft and hard X-ray bands with three models starting with a phenomenological model for comparison with past work. The models get progressively complex with a thermal Comptonization model \texttt{compps} \citep{Poutanen96} and a hybrid Comptonization model \texttt{eqpair} \citep{Coppi99}. We have chosen these models for three reasons: 1. to be able to infer physical characteristics of the Compton scattering in the medium, 2. to search for additional power-law components that may arise from a compact jet, and 3. to make a comparison with past work which uses the same models.

The composite \texttt{diskbb} + \texttt{power law} model provided acceptable fits but over-estimated the counts above 100 keV in all spectra. Adding a high-energy cut off (\texttt{highecut}) to the model was significantly preferred in the F-test (P < 10$^{-12}$). The folding energies were typically above 100 keV (within 1.6$\sigma$ error). We calculated 0.5-350 keV fluxes of the source using the best-fit models. The results of the spectral analysis are tabulated in Table~\ref{tab:highepow}.

For the \texttt{compps} model \citep{Poutanen96} we used an external \texttt{diskbb} component in the analysis of the first four data as justified in Section~\ref{sect:states}. We chose a spherical geometry for the corona, assuming the seed photons to be the soft photons originating from the inner edge of the multicolor disc blackbody, and left the electron temperature and the optical depth free but fixed the rest of the model parameters at their default values. We considered both the thermal and the hybrid plasma cases. For the former, the best-fit parameters as well as the fit statistics are listed in Table~\ref{tab:diskbbcompps}. Compton reflection was ignored in all fits as it never exceeded 0.05 (3$\sigma$ upper limit). For the latter, the non-thermal electrons were injected with a power-law of $\Gamma_p$ = 2.5 between Lorentz factors $\gamma_{min}$ = 1.3 and $\gamma_{max}$ = 1000. This resulted in electron temperature pegging at its lowest allowed value of 20 keV and the optical depth pegging at its maximum allowed value of 3. Furthermore, leaving the $\Gamma_p$ free did not improve the results. We, therefore, conclude that our hybrid plasma fits with the \texttt{compps} model was inadequate to explain the data as some parameters are not physically constrained.

For the \texttt{eqpair} model we again used an external \texttt{diskbb} component in the analysis of the first four data sets. We started with the purely thermal Comptonization case by fixing the $l_{nt}/l_h$=0. We assumed the seed photon temperature to be that in \texttt{diskpn} model, fixed the soft-photon compactness to $l_s$ = 1, and the reflection strength $R$ = 0. We left the seed photon temperature $T_{seed}$, the hard-to-soft compactness $l_h/l_s$, the optical depth ($\tau$) free but the rest of the parameters fixed at their default values. We also performed fits for the hybrid thermal/non-thermal Comptonization cases by setting the parameters of the electron distribution as in the hybrid \texttt{compps} fits and leaving the $l_{nt}/l_h$ parameter free. The results for both the thermal and hybrid Comptonization fits are listed in Table~\ref{tab:diskbbeqpair}. We also show two representative broadband spectra of ~\SO\ together with the $\Delta\chi$ values of all the fit models in Figure~\ref{fig:broadband}.

\subsubsection{Broadband SEDs}\label{sect:broadband}
We constructed broadband SEDs for three epochs from the main outburst decay and one from the one from the mini-outburst. The \textit{Swift}/UVOT data were transformed to the flux units during the photometric extraction (see Section~\ref{sect:Swiftred}), and the SMARTS optical/NIR and the TUG optical data were done with the zero points in \cite{Bessell98}. All the UOIR data were dereddened using a color excess of E(B-V) = 0.16 $\pm$ 0.05 (see Section~\ref{sect:xray}) transformed to wavelength dependent extinction values with the reddening curve in \cite{Fitzpatrick07} assuming R$_V$ = 3.1. The errors on the flux densities include uncertainties on both the photometry and the extinction correction.

\begin{figure*}
 \begin{multicols}{2}
  \includegraphics[width=\columnwidth]{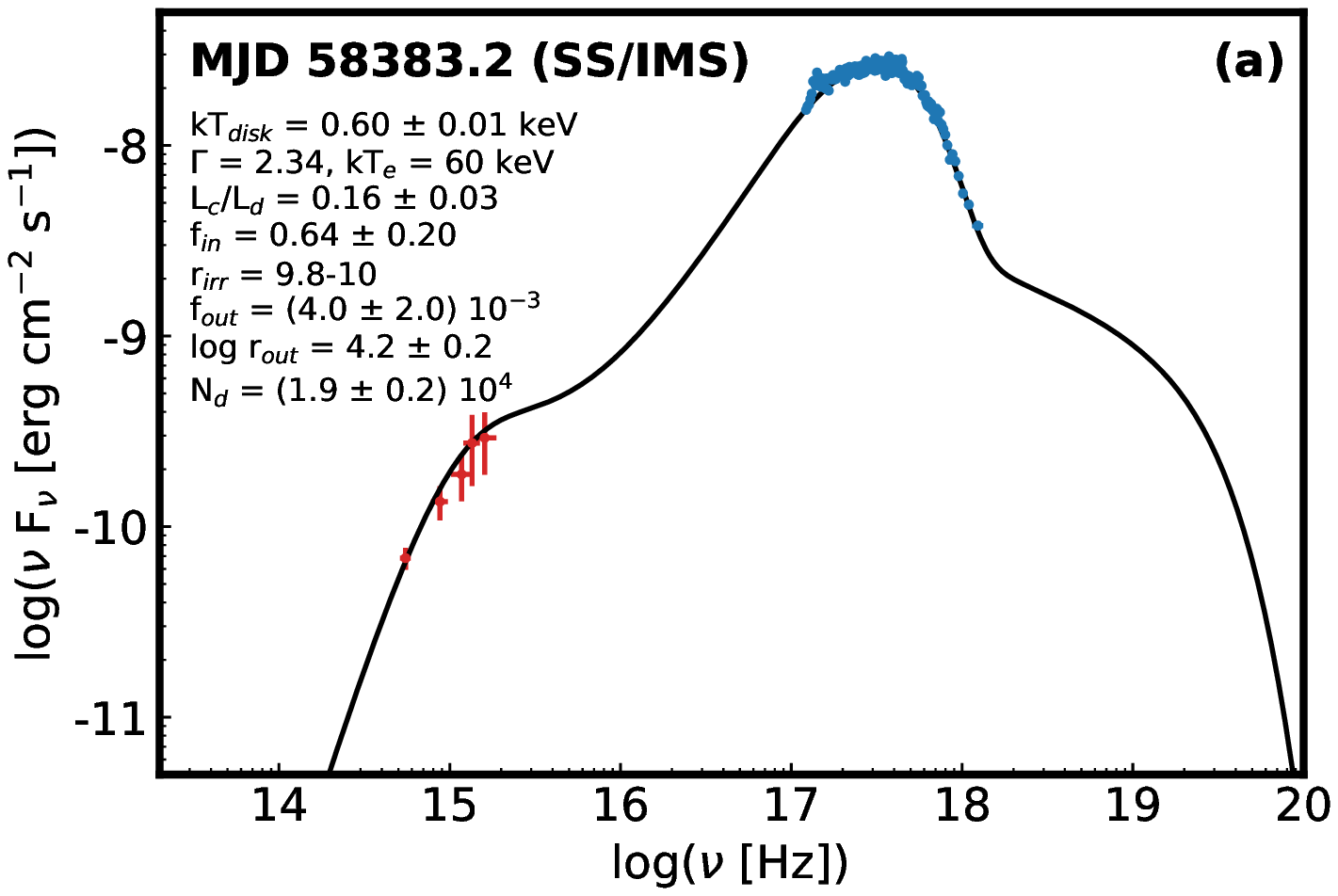}\par
  \includegraphics[width=\columnwidth]{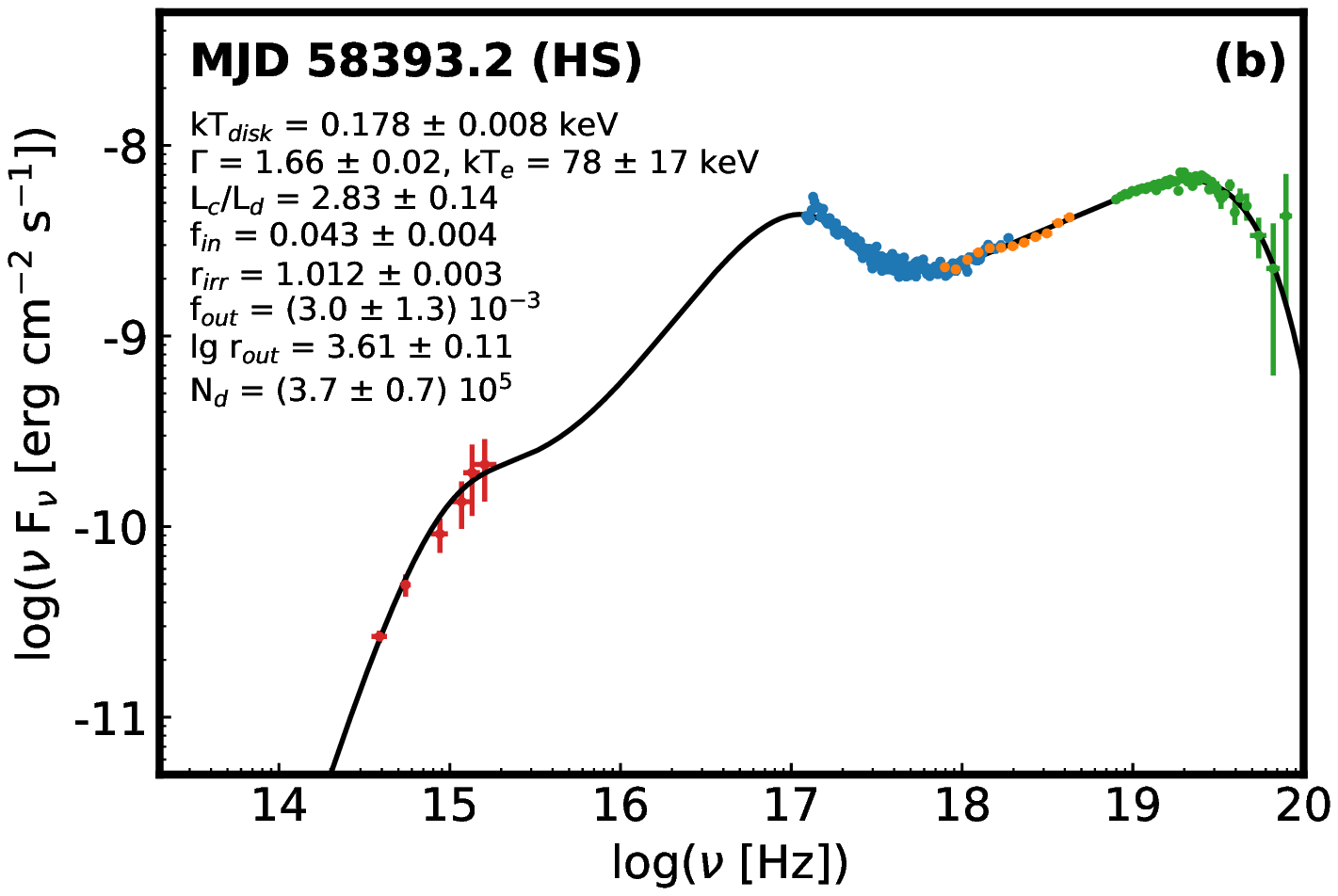}\par
  \end{multicols}
 \begin{multicols}{2}
  \includegraphics[width=\columnwidth]{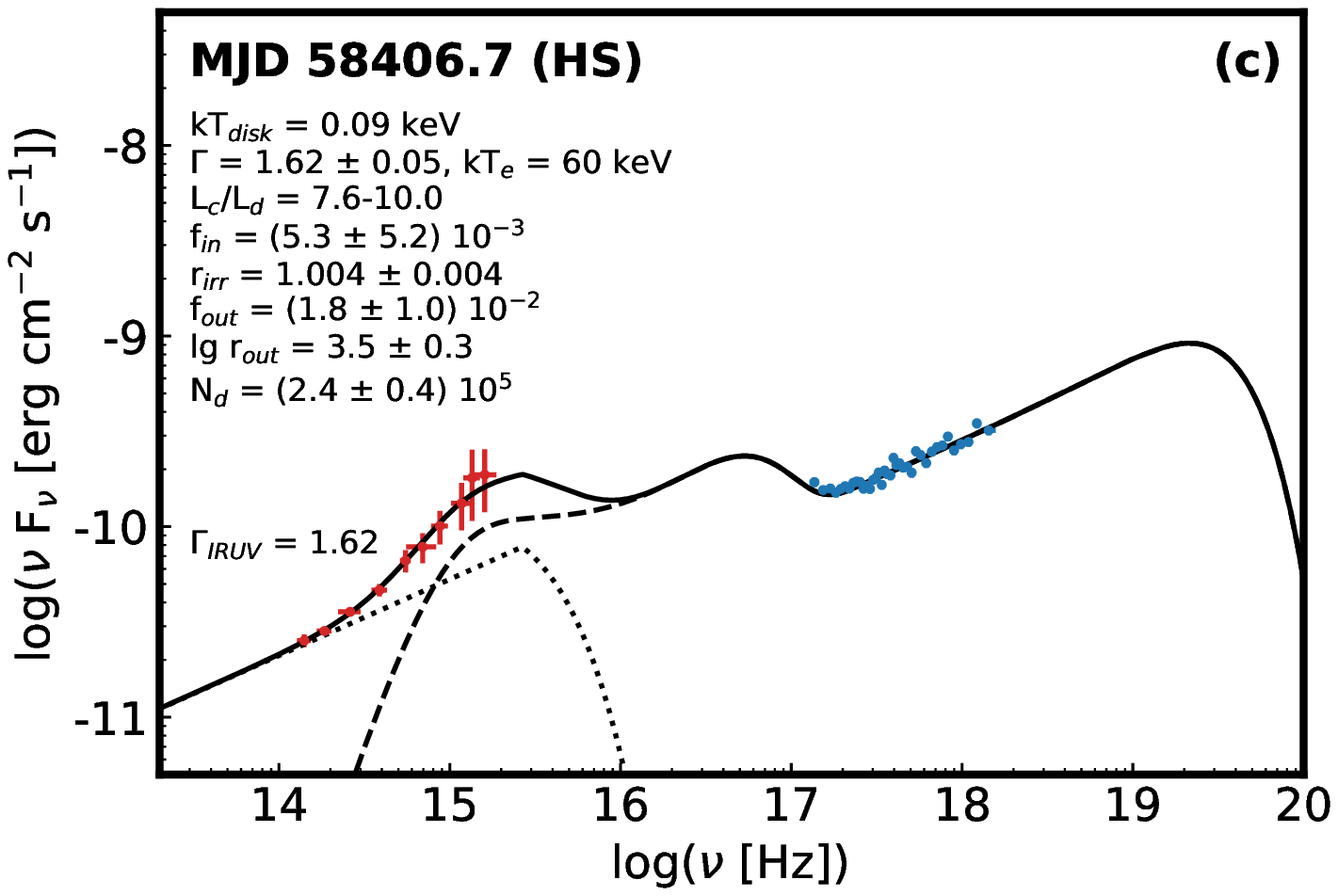} \par
  \includegraphics[width=\columnwidth]{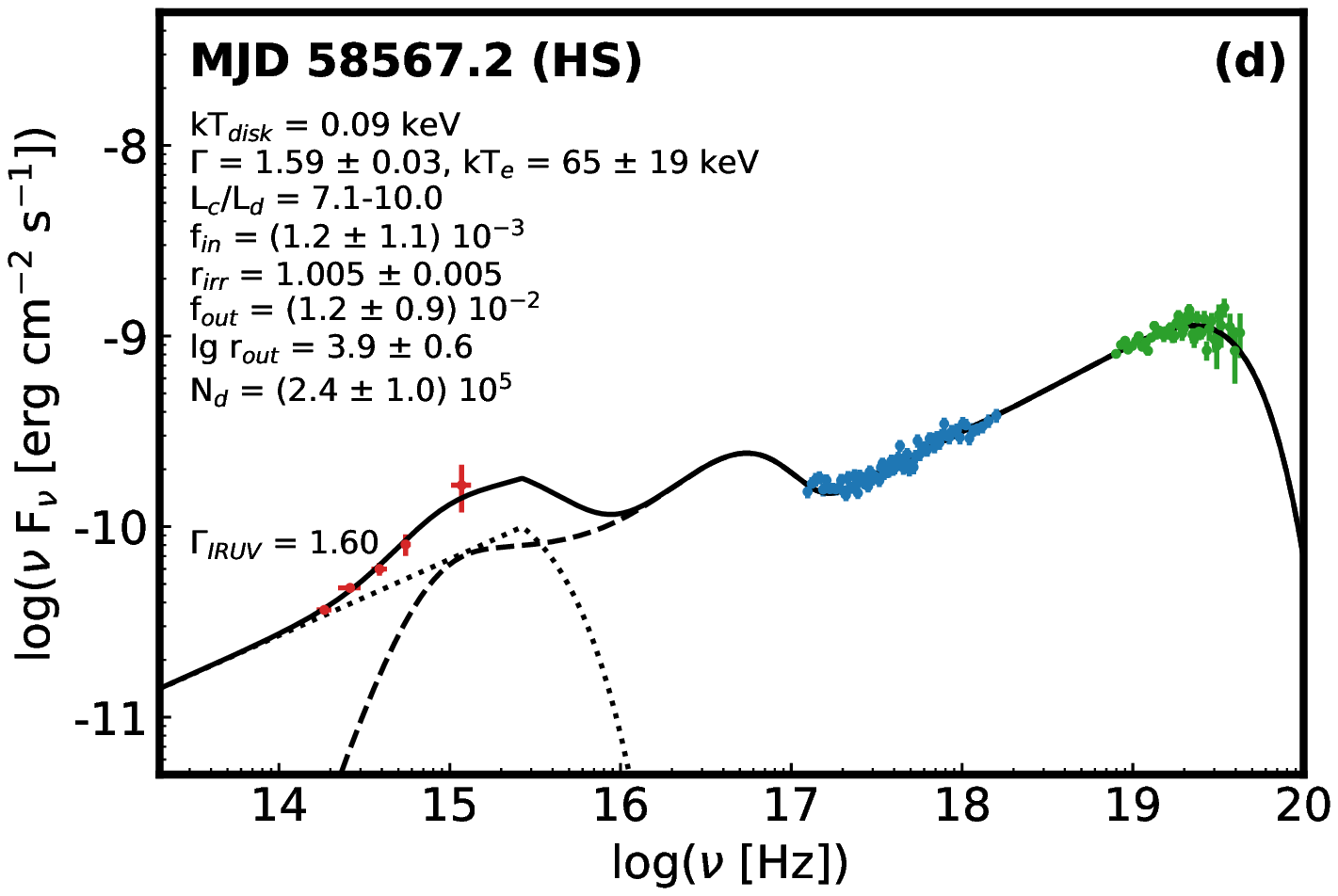} \par
  \end{multicols}
 \caption{Unfolded broadband SEDs of MAXI~J1820$+$070 from (a) the soft/intermediate state, (b)-(c) the hard states of the main outburst decay, and (c) the hard mini-outburst. The red data points correspond to the UOIR observations from UVOT, TUG and SMARTS. XRT, JEM-X and ISGRI observations are given in blue, orange and green data points, . The SEDs in panel (a)-(b) were fit with the \texttt{diskir} model whereas the ones in panel (c)-(d) with the \texttt{diskir} + \texttt{power law} composite model (see Section~\ref{sect:broadband})}.
 \label{fig:SEDs}
\end{figure*}

In order to investigate the spectral components contributing to these bands we performed SED fittings. The best-fit results are shown in Figure~\ref{fig:SEDs}. The first two SEDs, which were taken before the rebrightening in UOIR band (on MJD 58383.2 and 58393.2, ), were in good agreement with the irradiation model \texttt{diskir} \citep{Gierlinski08, Gierlinski09}. On the other hand, the SEDs during the rebrightening (MJD 58406.7) and the mini-outburst (MJD 58567.2), produced strong residuals in the NIR-UV region. To improve these last two fits, we added a power-law of $\Gamma$=1.6 with a sharp high-energy cut off at 4 $\times$ 10$^{15}$ Hz to the model, and obtained acceptable results. The photon index we used here corresponds to spectral index of $\alpha$ = -0.6, a value that is provided by the standard particle acceleration theory \citep{Bednarz98, Kirk00}. Note that a similar approach was also taken by \cite{Shidatsu18} to explain the SED during the hard state rise (around MJD 58201).

\section{Discussion} \label{sect:discussion}

\subsection{Multiwavelength evolution }
Our results show that MAXI~J1820$+$070 exhibited first an increase in the hard X-rays as a signature of the soft-to-intermediate state transition and then a rebrightening concurrently in all of the UOIR band in the hard state. The rebrightening in UOIR band was observed $\sim$ 7 days after the hard state transition. This multiwavelength behaviour is frequently observed in other GBHTs \citep{Kalemci05, Kalemci13, Dincer12, Baglio18}. 

In GBHTs, the origin of the UOIR emission depends on the X-ray state of the source. In the soft and intermediate states, the observed emission is produced by the irradiated outer accretion disc \citep{Russell06, Rykoff07, Gierlinski09} whereas in the hard state, starting from the onset of the rebrightening, at least one additional component contributes to the emission in these bands. This is also true for MAXI~J1820$+$070 as its SEDs until the onset of the rebrightening (Figure~\ref{fig:SEDs}(a) and (b)) are consistent with the irradiated outer accretion disc, while both the one on the top of the rebrightening and the hard mini-outburst (Figure~\ref{fig:SEDs}(c) and (d)) are only able to provide good fits with an additional power-law component extending from the radio to the UV band (see Section~\ref{sect:broadband}). 
\begin{figure}
\centering
\includegraphics[width=\columnwidth]{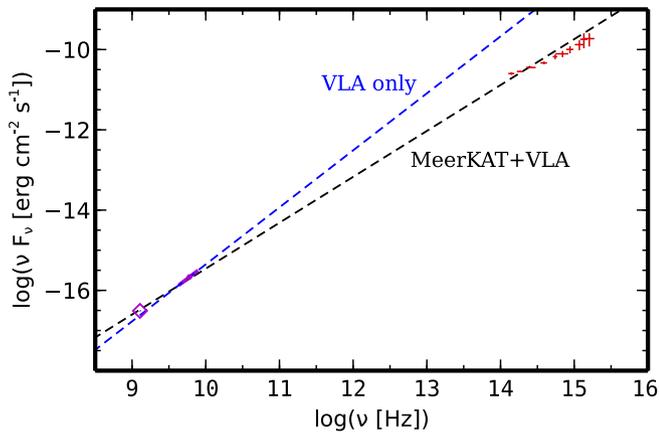}
\caption{Unfolded radio-UOIR SED for MAXI J1820+070. MeerKAT (MJD 58405.67) is shown with a diamond, VLA (MJD 58405.90) is shown with a solid line (magenta) between 4.5 - 7.5 GHz with the measured radio spectral index, and the red points are the SMARTS and UVOT fluxes also shown in Fig.~\ref{fig:SEDs} (c). The blue dashed line is a power-law with the VLA spectral index, and the black dashed line is obtained by fitting a power-law to the entire data set.}
\label{fig:SED-radio}
\end{figure}
We looked into this additional power-law component in more detail with the contemporaneous MeerKAT/VLA/SMARTS/UVOT data around MJD 58406.7. In Fig.~\ref{fig:SEDs} (c) we showed an overall fit to the SED with a model consistent with other observations in the same figure that we do not have the radio data for. However, when the available radio data are included, as shown in Fig.~\ref{fig:SED-radio}, more details emerge. The radio spectrum is slightly inverted, and a single power-law fit to joint MeerKAT/VLA/SMARTS/UVOT data yields a spectral index $\alpha$=0.14. While the single power-law connects radio to UOIR data as seen in Fig.~\ref{fig:SED-radio} with the black dashed lines, the $\chi^{2}$ is 149.5 for 10 degrees of freedom. The poor fit is a result of the small errors in the radio and infrared flux measurements. We note that since the data are not exactly simultaneous, and both the infrared and radio emissions are variable (see Figs. \ref{fig:uvoptir} and \ref{fig:multievol}), the systematic errors are possibly much larger than the measurement errors. The blue dashed lines show the extended power-law emission with $\alpha$=0.42 based on the VLA data only \citep[see][for radio spectral indices based on VLA observations]{Shaw21}. The infrared to UV frequency band is at the intersection of several emission components operating at the same time, including possibly more than one jet related component \citep[e.g.][]{Rodi21}, and groups with realistic jet models are welcome to a more complex fit to this data set with the data becoming available. While it may not be exactly clear where the spectral break occurs, adding the radio data strengthens the argument for the presence of a jet related power-law component in this region.

For the hard state observation during the rise of the outburst on March 24, 2018, \cite{Shidatsu18} claimed that a similar additional power-law component had a jet origin. \cite{Zdziarski2022} and \cite{Tetarenko21} were both able to show that the broadband SED in the outburst rise with the radio data can be fitted with a jet model. \cite{Tetarenko21} also used multiwavelength timing studies to deduce additional jet parameters. While the usual interpretation for such a component extending to the near infrared and beyond is the synchrotron emission from jets \citep[][and references therein]{Buxton04, Kalemci05, Russell13}, synchrotron emission from a hot accretion flow is also suggested \citep[][and references therein]{Veledina13}. An optical polarimetry study of this source also proved to be inconclusive to distinguish the origin of this additional component \citep{Veledina19}; however SED fitting with realistic jet models or hot accretion flow model could provide further information. 

\cite{Bright18} detected radio emission from the source with a flux density of $\sim$0.9 mJy at 15.5 GHz on MJD 58385.6 with AMI-LA, $\sim$ 12 days prior to the rebrightening. This is consistent with the jet revival in other sources during their soft-to-hard state transition \citep{MillerJones12,Corbel13,Kalemci13,Russell13,RussellTD14}. Moreover, the radio brightness measured by MeerKAT increased between MJD 58389 and MJD 58397, while the source was fading in all other wavelengths, consistent with a decay rate of ~21 days \citep{Bright20}. For simultaneous and contemporaneous MeerKAT and VLA data, the radio spectrum is inverted, indicating a compact jet. A comparison with the behaviour of GX 339-4 during its 2010-2011 decay can be made here. For GX 339-4, the radio brightening has started around 10 days before the onset of the NIR rebrightening and the first detection of the compact jet. The radio observations before the NIR rebrightening indicated an optically thin emission and became optically thick as the NIR flux increased. The radio flux stayed almost at constant levels for around 10 more days, until the peak of the NIR emission. Afterwards, both NIR and radio fluxes decreased. This has been interpreted as the NIR emission is mostly coming from the compact radio jet during the rebrightening, and the radio spectral index evolves from an optically thin to optically thick emission at the onset of the NIR rebrightening \citep{Corbel13}. For \SO, the radio flux increases before the UOIR brightening in similar timescales to those of GX 339-4 (though there is no radio spectral index information before MJD ~58398). But the striking difference with the behaviour of GX~339-4 is that the radio flux in \SO\ is already decreasing between MJD~58398 and MJD~58402, while the optical and UV fluxes are still increasing. Similar behaviour is also observed in MAXI~J1836$-$194 \citep{Russell13, RussellTD14} for which the radio emission is already fading with the X-rays, while the optical-IR flux is still rising. While the relative timing of radio flux and the UOIR flux increase is consistent with the standard jet formation model as the jet reestablishes itself brightening first at longer wavelengths before UOIR \citep{MillerJones12x, Corbel13}, the subsequent relative evolution changes from source to source.

\subsection{High energy behaviour}
The high energy behavior can be discussed in terms of the phenomenological and Comptonization model fits to the XRT+JEM-X+ISGRI joint data (see \S\ref{sect:spectra} for all the fit results). For the phenomenological model, the first four observations during the outburst decay provided typical hard state folding energies as given in Table~\ref{tab:highepow}. In the model we used, the folding energies are close to the peak energy output of the Comptonization and therefore corresponds to 3$kT_{e}$ where $k$ is the Boltzman constant and $T_{e}$ is the electron thermal temperature. These temperatures are consistent with the Comptonization electron temperatures with the \texttt{compps} fits, which stayed between 60 keV and 80 keV as given in Table~\ref{tab:diskbbcompps}. 
Figure~\ref{fig:multievol} shows that these four observations between MJD~58393-58398 occurred at a time when the soft X-ray spectral index is hardening towards its minimum value, right before the rebrightening in UOIR, and during the time that the radio flux is increasing. While there is no apparent evolution in the Comptonization parameters (optical depth $\tau$ and $T_{e}$ of the Comptonizing corona) in the \texttt{compps} model fits, there is an apparent drop in the disk temperature. This can also be seen in soft X-ray fits (see Fig.~\ref{fig:broadband} and Table~\ref{tab:SwiftonlyNhfixed}).

The \texttt{eqpair} model allows changing a variety of Comptonization and geometrical parameters, and we followed the recommendations in \cite{Coppi99} to restrict the parameters as described in Section~\ref{sect:observations}. In this model, it is hard to determine the Comptonization parameters hard-to-soft compactness $l_{h}/l_{s}$, the Thomson scattering depth $\tau_{p}$ and the reflection parameters independently if the quality of the data is not great. Therefore, we only used the \texttt{eqpair} model to assess the possibility of the hybrid electron energy distribution and also to compare with the past results from other sources. Indeed, the \texttt{eqpair} fits indicate that a highly non-thermal model is preferred over pure thermal Comptonization for these observations whereas for the last observation, taken during the mini-outburst, a thermal distribution of electron energies is preferred (see Table~\ref{tab:diskbbeqpair}). Moreover, none of the models, phenomenological or Comptonization, has been improved by adding extra power-law component sometimes associated with the jet emission.

The outburst rise of this source from the hard state to the intermediate states beyond the $NuStar$ band ($>$ 70 keV) has been studied with $INTEGRAL$, $MAXI$, and $HXMT$. The region of interest to compare with our data would be $\pm$5 days around MJD~58200. During this time the source is evolving from a very hard state to an intermediate state. Both $MAXI$ \citep{Shidatsu18} and $HXMT$ \citep{You21} data fits around that time resulted in Comptonization electron temperatures of 40-80 keV, similar to the temperatures we obtained during the decay. We note that the fit models are different, therefore this is not a direct comparison. 

The most detailed high energy study of the source has recently been presented in \cite{Zdziarski21} for 2 epochs in outburst rise with spectra extending to MeV range using data from $NuStar$, ISGRI and SPI on $INTEGRAL$ indicating the presence of a hybrid distribution of electron energies in the Comptonization process together with thermal Comptonization. Their approach with two Comptonization model does not necessarily indicate an additional component from a jet.  We note that this comparison with the outburst rise could only be done indirectly due to two important reasons. Thanks to the quality of outburst rise data, together with high spectral resolution of $NuSTAR$ and the extended energy range to 2 MeV with the INTEGRAL SPI and PiCSIT, \cite{Zdziarski21} were able to obtain strong constraints on accretion geometry and associated physical parameters. Moreover, they used different Comptonization models (\texttt{reflkerr} and \texttt{reflkerr.bb}, see \citealt{Zdziarski21} and references therein) more appropriate for the data they use. While the thermal Comptonization part still uses \texttt{comppps} in the background, the non-thermal part is calculated self-consistently in a different manner than \texttt{eqpair}. We should note that although our data also prefers non-thermal distribution of electron energies, the quality of the spectra prohibits us constraining physical parameters any further. 

We could also compare our results with the other GBHTs studied with $INTEGRAL$ during the outburst decays. With our $INTEGRAL$ observing program of GBHTs during the decay we have observed XTE~J1752$-$223 \citep{Chun13} and SWIFT~J1745$-$26 \citep{Kalemci14}. For both of these sources, the $INTEGRAL$ observations took place after the source reached its hardest levels. In a sense, the observations of \SO\ is complementary to the already present hard state observations, as the current data set includes the evolution of the system towards the hard state. For the simple \texttt{highecut} fits done in the same way for all observations during the decay, one can observe that both XTE~J1752$-$223 and SWIFT~J1745$-$26 have higher folding energies compared to \SO. In parallel to this, when fitted with \texttt{compps}, we observe that XTE~J1752$-$223 has a higher temperature in the Comptonizing medium, whereas for SWIFT~J1745$-$26, it either has a higher temperature, or higher optical depth. These results are not surprising at all, for ~\SO\ both the electron temperature and density is lower compared to those of other sources already deep in the hard state. On the other hand, our observations do not allow us to track the evolution of the Comptonization parameters with the given errors, the only significant evolution we can observe is in the decreasing disc flux and the temperature. Interestingly, the \texttt{eqpair} fits in SWIFT~J1745$-$26 deep in the hard state does not necessarily require non-thermal electron distribution. For XTE~J1752$-$223, the thermal Comptonization is a good fit as well. The lack of data beyond MJD~58400 makes it hard to discuss if there is an evolution from hybrid to thermal distribution in outburst decays, this is something that can only be tested with observations catching both the transition and hard state observations in one source. 

Finally, we can compare behaviour of MAXI J1820$+$070 during the mini-outburst. All three sources we investigated with \textit{INTEGRAL} during the outburst decay \citep{Kalemci14,Chun13} and V404 Cyg showed such late brightening events after the main outburst \citep{Kajava18}. For XTE~J1752$-$223, SWIFT~J1745$-$26, and \SO, the episodes started $\sim$60, $\sim$100, and $\sim$165 days after the soft-to-hard state transition in the main outburst decay, respectively. In all three cases, the increase in the hard and the soft X-ray flux is accompanied with an increase in OIR band and lasted for around 40 days. Both SWIFT~J1745$-$26 and \SO\ spectral fits indicated that thermal Comptonization is enough to explain the hard X-ray spectra. Similarly, the June 2015 mini-outburst of V404 Cyg occurred $\sim$150 days after the main outburst, and showed X-ray and radio flares, and rich P-Cygni profiles indicative of winds in the hard state \citep{Munoz17}. The spectral analysis indicated hard state with behavior similar to the beginning of the regular outburst \citep{Kajava18}.

For \SO, the apparent lack of any delay between UOIR emission and hard X-rays \citep[which is also observed in XTE J1752$-$223][]{Chun13} and spectral indices starting from the hard levels compared to the softer low flux observations \citep{Shaw21}, all indicate that the mini-outburst behaves like a start of a new outburst consistent with additional mass accretion on to the compact object perhaps due to the change of the nature of emission during the state transition to the hard state. However, for the four cases investigated with INTEGRAL, the mini-outbursts did not make a transition to the soft state. This is the case for most sources, but for MAXI~J1535$-$571, the multiple outbursts seemed to go into softer states as indicated through their HIDs \citep{Cuneo20}. We note that MAXI~ J1535$-$571 mini-outbursts, while separated by days, became quite bright, and the source never entered the quiescence. The origin and behavior of the increased flux levels after the main outburst decays could be different from source to source showing the importance of following the entire outburst with multiwavelength coverage.
\section{SUMMARY}\label{sect:sum}
In this work, we have studied the spectral properties of MAXI J1820+070 during the decay of the 2018 outburst and the subsequent mini-outburst in March 2019. Using quasi-simultaneous data from INTEGRAL, Swift, SMARTS and TUG telescopes, we have investigated the evolution of spectral parameters from a multiwavelength perspective to understand the physical mechanisms governing the increase in flux levels detected at the final stage of the outburst. The main findings from our analysis can be summarised as follows:

(i) The source underwent a rebrightening event in hard X-rays and the UOIR band between MJD ~58380-58440, close to the end of the main outburst. A few days after the radio detection, it transitioned from the intermediate-to-hard state on MJD 58390. The rebrightening started in the hard X-rays first and was followed by the UOIR bands. We have found the onset of the rebrightening in the V-band on MJD 58397.5, a week after the hard state transition, a well-known multiwavelength behaviour seen in other GBHTs observed at the outburst decay.

(ii) We showed that the source stayed in the hard state during its $\sim$40-day long mini-outburst. In contrast to the rebrightening, it started ~165-days after the soft-to-hard state transition at the main outburst decay and showed a similar trend in its multiwavelength light curves. We compared the spectral properties and the measured time scales of mini-outburst with other GBHTs investigated with INTEGRAL and concluded that it behaved like a new outburst.

(iii) We have performed broadband spectral fitting on the joint XRT+JEM-X+ISGRI data using phenomenological and Comptonization models. A non-thermal electron energy distribution is preferred over a pure thermal distribution for the rebrightening phase, however, the mini-outburst case can be fitted with a pure thermal Comptonization model. Furthermore, none of the models has been improved by adding an extra power-law component (possibly related to the jet) in X-rays in contrast to models presented in the outburst rise. We stress that the quality of the data in the outburst rise is significantly better than that of outburst decay.

(iv) The SEDs until the onset of the rebrightening are consistent with the irradiated outer accretion disc, while the ones close to the UOIR and mini-outburst peaks provide good fits only with an additional power-law component in the NIR-UV band. We looked into the detail of this power-law component for the data around MJD 58406.7 by adding the radio observations to the fits and discussed that it might have a jet origin though we could not constrain the exact location of the spectral break.

\section*{Acknowledgements}
We thank the anonymous referee for constructive comments that improved the paper. M\"{O}A acknowledges support from the Royal Society through the Newton International Fellowship programme. EK acknowledges T\"{U}B\.{I}TAK Project 115F488. DA acknowledges support from the Royal Society. We thank the TUG director Dr. S. \"{O}zdemir, for approving our DDT observations and TUBITAK for a partial support in using T100 telescope with project number 1423. We also thank E. Sipahi, M. Kaplan, \"{O}. Ba\c{s}t\"{u}rk, M. Acar, D. S\"{u}rgit, \.{I}. Nas{\i}ro\u{g}lu, A. Ivantsov, M. Yard{\i}mc{\i} and H. Eseno\u{g}lu for providing time from their schedules to observe the source through the TUG-DDT/TOO programme. TD thanks Dr. P. B. Stetson for providing his DAOPHOT4 package. The authors also thank Jingyi Wang for sharing the Nicer/HID data of the source used in creation of Fig.~\ref{fig:HID}(b). This research has used data from the SMARTS 1.3-m telescope, which is operated as part of the SMARTS Consortium and MAXI data provided by RIKEN, JAXA and the MAXI team.

\section*{Data Availability}
The \textit{Swift} data underlying this article are publicly available in the HEASARC Data Archieve at \url{https://heasarc.gsfc.nasa.gov/}. The \textit{INTEGRAL} data with ObsIDs 15400050001 (PI: E. Kalemci) and 16700020001 can be reached at \url{https://www.isdc.unige.ch/integral/archive}. The MAXI/GSC data used in Fig.~\ref{fig:uvoptir} can be downloaded from \url{http://maxi.riken.jp/star_data/J1820+071/J1820+071.html}. The SMARTS data (PI: C. Bailyn) are available from \url{https://astroarchive.noao.edu/}. The TUG data (PI: T. Din\c{c}er) will be shared on reasonable request to the corresponding author. All the generated data underlying this study are available in the article and in its online supplementary material.



\bibliographystyle{mnras}
\bibliography{blackholebib} 



\section*{SUPPORTING INFORMATION}
Supplementary data are available at MNRAS online
\newpage
\appendix
\section{UOIR Monitoring Data from Swift/UVOT, TUG and SMARTS.}
\begin{table*}
\begin{threeparttable}
\caption{\textit{Swift}/UVOT measurements.}
\begin{tabular}{cccccccccc}
\toprule 
Date & \textit{V} mag  & Date & \textit{U} mag & Date & \textit{M2} mag & Date & \textit{W1} mag& Date & \textit{W2} mag\\
\midrule
383.1560& 14.167 $\pm$ 0.027& 378.9005& 12.929 $\pm$ 0.024& 378.8955& 12.790 $\pm$ 0.032& 378.8986& 12.683 $\pm$ 0.032& 378.9030& 12.626 $\pm$ 0.032\\
386.0776& 14.074 $\pm$ 0.037& 383.1625& 13.204 $\pm$ 0.024& 383.1581& 13.013 $\pm$ 0.033& 383.1608& 12.945 $\pm$ 0.033& 383.1534& 12.879 $\pm$ 0.032\\
387.6087& 14.073 $\pm$ 0.027& 386.0809& 13.070 $\pm$ 0.027& 386.0786& 12.990 $\pm$ 0.036& 386.0799& 12.849 $\pm$ 0.036& 386.0763& 12.819 $\pm$ 0.033\\
388.9240& 14.262 $\pm$ 0.029& 387.6148& 13.185 $\pm$ 0.025& 387.6107& 13.095 $\pm$ 0.034& 387.6132& 12.948 $\pm$ 0.033& 387.6062& 12.943 $\pm$ 0.032\\
390.3235& 14.374 $\pm$ 0.029& 388.9300& 13.360 $\pm$ 0.026& 388.9260& 13.241 $\pm$ 0.034& 388.9285& 13.152 $\pm$ 0.034& 388.9215& 13.086 $\pm$ 0.032\\
391.4571& 14.472 $\pm$ 0.030& 390.3305& 13.444 $\pm$ 0.024& 390.3258& 13.328 $\pm$ 0.034& 390.3287& 13.220 $\pm$ 0.034& 390.3207& 13.158 $\pm$ 0.032\\
392.5861& 14.495 $\pm$ 0.031& 391.4516& 13.624 $\pm$ 0.026& 391.4597& 13.411 $\pm$ 0.033& 391.4499& 13.289 $\pm$ 0.034& 391.4543& 13.197 $\pm$ 0.032\\
393.2410& 14.517 $\pm$ 0.032& 392.5808& 13.650 $\pm$ 0.027& 392.5888& 13.405 $\pm$ 0.033& 392.5792& 13.326 $\pm$ 0.034& 392.5834& 13.259 $\pm$ 0.032\\
394.7634& 14.696 $\pm$ 0.061& 393.2472& 13.627 $\pm$ 0.025& 393.2430& 13.409 $\pm$ 0.035& 393.2455& 13.306 $\pm$ 0.034& 393.2385& 13.238 $\pm$ 0.032\\
395.4312& 14.656 $\pm$ 0.033& 394.7657& 13.706 $\pm$ 0.030& 394.7641& 13.458 $\pm$ 0.043& 394.7650& 13.451 $\pm$ 0.043& 394.7625& 13.318 $\pm$ 0.037\\
397.0990& 14.649 $\pm$ 0.033& 395.4377& 13.764 $\pm$ 0.027& 395.4333& 13.497 $\pm$ 0.035& 395.4360& 13.465 $\pm$ 0.034& 395.4285& 13.341 $\pm$ 0.032\\
400.0145& 14.240 $\pm$ 0.037& 397.1056& 13.754 $\pm$ 0.026& 397.1012& 13.623 $\pm$ 0.035& 397.1038& 13.498 $\pm$ 0.035& 397.0964& 13.414 $\pm$ 0.032\\
402.2814& 14.027 $\pm$ 0.033& 400.0180& 13.586 $\pm$ 0.047& 400.0157& 13.489 $\pm$ 0.038& 400.0173& 13.351 $\pm$ 0.037& 400.0130& 13.298 $\pm$ 0.034\\
404.2669& 14.016 $\pm$ 0.025& 402.2851& 13.327 $\pm$ 0.029& 402.2826& 13.326 $\pm$ 0.037& 402.2842& 13.133 $\pm$ 0.036& 402.2798& 13.209 $\pm$ 0.034\\
406.7369& 14.203 $\pm$ 0.029& 404.2733& 13.367 $\pm$ 0.030& 404.2691& 13.360 $\pm$ 0.034& 404.2719& 13.198 $\pm$ 0.034& 404.2641& 13.262 $\pm$ 0.032\\
408.2500& 14.141 $\pm$ 0.028& 406.7427& 13.535 $\pm$ 0.029& 406.7389& 13.476 $\pm$ 0.035& 406.7413& 13.325 $\pm$ 0.034& 406.7345& 13.373 $\pm$ 0.033\\
410.3100& 14.276 $\pm$ 0.026& 408.2563& 13.512 $\pm$ 0.025& 408.2521& 13.477 $\pm$ 0.035& 408.2546& 13.281 $\pm$ 0.034& 408.2475& 13.353 $\pm$ 0.032\\
412.0373& 14.278 $\pm$ 0.023& 410.3185& 13.505 $\pm$ 0.023& 410.3127& 13.531 $\pm$ 0.034& 410.3161& 13.368 $\pm$ 0.033& 410.3066& 13.393 $\pm$ 0.032\\
417.4849& 14.570 $\pm$ 0.026& 412.1036& 13.608 $\pm$ 0.026& 412.0404& 13.562 $\pm$ 0.034& 412.1023& 13.460 $\pm$ 0.038& 412.0331& 13.423 $\pm$ 0.032\\
419.6093& 14.507 $\pm$ 0.041& 417.4949& 13.818 $\pm$ 0.024& 417.4882& 13.831 $\pm$ 0.034& 417.4923& 13.673 $\pm$ 0.033& 417.4808& 13.647 $\pm$ 0.032\\
424.6648& 14.782 $\pm$ 0.030& 419.6129& 13.972 $\pm$ 0.032& 419.6105& 13.836 $\pm$ 0.041& 419.6120& 13.640 $\pm$ 0.039& 419.6078& 13.731 $\pm$ 0.036\\
425.4588& 14.943 $\pm$ 0.042& 424.6736& 14.189 $\pm$ 0.073& 424.6680& 14.061 $\pm$ 0.035& 424.6719& 13.932 $\pm$ 0.034& 424.6609& 13.940 $\pm$ 0.033\\
426.6530& 14.793 $\pm$ 0.032& 425.4641& 14.126 $\pm$ 0.028& 425.4605& 14.138 $\pm$ 0.040& 425.4626& 13.903 $\pm$ 0.038& 425.4567& 13.998 $\pm$ 0.035\\
428.1772& 14.970 $\pm$ 0.036& 428.1847& 14.182 $\pm$ 0.026& 428.1796& 14.200 $\pm$ 0.037& 428.1827& 14.096 $\pm$ 0.036& 426.6489& 13.992 $\pm$ 0.032\\
430.7013& 15.165 $\pm$ 0.048& 430.7065& 14.454 $\pm$ 0.032& 430.7030& 14.309 $\pm$ 0.041& 430.7051& 14.207 $\pm$ 0.040& 428.1741& 14.072 $\pm$ 0.034\\
434.3607& 15.328 $\pm$ 0.040& 434.3690& 14.474 $\pm$ 0.033& 434.3636& 14.600 $\pm$ 0.039& 434.3672& 14.445 $\pm$ 0.037& 430.6993& 14.275 $\pm$ 0.036\\
436.5399& 15.417 $\pm$ 0.043& 436.5481& 14.720 $\pm$ 0.033& 436.5427& 14.782 $\pm$ 0.041& 436.5462& 14.587 $\pm$ 0.038& 434.3571& 14.576 $\pm$ 0.035\\
438.5992& 15.704 $\pm$ 0.046& 438.6096& 14.782 $\pm$ 0.029& 438.6027& 15.007 $\pm$ 0.040& 438.6070& 14.630 $\pm$ 0.037& 436.5365& 14.577 $\pm$ 0.035\\
440.1928& 15.756 $\pm$ 0.047& 440.2033& 15.125 $\pm$ 0.031& 440.1963& 15.167 $\pm$ 0.042& 440.2006& 14.931 $\pm$ 0.039& 438.5950& 14.899 $\pm$ 0.035\\
559.1445& 14.226 $\pm$ 0.038& 559.1482& 13.497 $\pm$ 0.028& 558.8623& 13.541 $\pm$ 0.031& 555.8067& 14.042 $\pm$ 0.031& 440.1884& 14.936 $\pm$ 0.035\\
566.2450& 13.876 $\pm$ 0.024& 564.2554& 13.311 $\pm$ 0.021& 559.1457& 13.478 $\pm$ 0.039& 559.1472& 13.273 $\pm$ 0.037& 557.0690& 13.893 $\pm$ 0.032\\
$-$& $-$& 566.2516& 13.310 $\pm$ 0.026& 562.1250& 13.170 $\pm$ 0.031& 563.1209& 12.994 $\pm$ 0.031& 559.1431& 13.384 $\pm$ 0.034\\
$-$& $-$& 572.1505& 13.448 $\pm$ 0.021& 566.2472& 13.197 $\pm$ 0.034& 566.2500& 12.965 $\pm$ 0.033& 561.8594& 13.061 $\pm$ 0.031\\
$-$& $-$& 576.0631& 13.520 $\pm$ 0.021& 571.0312& 13.311 $\pm$ 0.032& 567.1729& 13.096 $\pm$ 0.031& 565.1816& 13.248 $\pm$ 0.031\\
$-$& $-$&$-$& $-$& 574.3344& 13.360 $\pm$ 0.031& 575.1414& 13.339 $\pm$ 0.031& 566.2423& 13.096 $\pm$ 0.032\\
$-$& $-$& $-$& $-$& $-$& $-$& $-$& $-$& 573.3455& 13.230 $\pm$ 0.031\\
$-$& $-$& $-$& $-$& $-$& $-$& $-$& $-$& 577.0589& 13.419 $\pm$ 0.031\\
\bottomrule
\end{tabular}
\begin{tablenotes}
\small
\item Note: Observation time is given as Date = MJD - 58000 in days.
\end{tablenotes}
\end{threeparttable}
\label{tab:uvot}
\end{table*}
\begin{table*}
\begin{threeparttable}
\caption{Measurements of optical and NIR data from SMARTS observations. A full version of this table is available online.}
\label{tab:smarts}
\begin{tabular}{cccccc}
\toprule 
Date & \textit{B} mag  & Date & \textit{V} mag & Date & \textit{I} mag \\
\midrule
406.0396& 14.513 $\pm$ 0.006& 406.0355& 14.195 $\pm$ 0.004& 406.0314& 13.508 $\pm$ 0.015\\
406.0406& 14.531 $\pm$ 0.005& 406.0365& 14.184 $\pm$ 0.005& 406.0324& 13.489 $\pm$ 0.015\\
406.0416& 14.550 $\pm$ 0.005& 406.0375& 14.184 $\pm$ 0.004& 406.0334& 13.502 $\pm$ 0.015\\
406.0426& 14.560 $\pm$ 0.005& 406.0385& 14.156 $\pm$ 0.004& 406.0344& 13.519 $\pm$ 0.016\\
407.0118& 14.507 $\pm$ 0.005& 407.0077& 14.199 $\pm$ 0.004& 407.0036& 13.548 $\pm$ 0.016\\
\vdots&\vdots&\vdots&\vdots&\vdots&\vdots\\
\midrule
&\textit{J} mag &&\textit{H} mag &&\textit{K} mag \\
\hline
406.0354& 12.724 $\pm$ 0.017& 406.0313& 12.092 $\pm$ 0.030& 407.0117& 11.412 $\pm$ 0.080\\
407.0076& 12.760 $\pm$ 0.017& 407.0035& 12.123 $\pm$ 0.031& 407.9972& 11.598 $\pm$ 0.080\\
407.9930& 12.810 $\pm$ 0.017& 407.9889& 12.227 $\pm$ 0.030& 409.0086& 11.626 $\pm$ 0.080\\
409.0045& 12.807 $\pm$ 0.017& 409.0004& 12.248 $\pm$ 0.031& $-$& $-$\\
410.9973& 12.901 $\pm$ 0.017& 410.9932& 12.366 $\pm$ 0.031& $-$& $-$\\
\vdots&\vdots&\vdots&\vdots& $-$& $-$\\
\bottomrule
\end{tabular}
\begin{tablenotes}
\item \footnotesize{Note: $^{*}$Date = MJD - 58000 in days}
\item \footnotesize{Note: Effective wavelengths ($\lambda_{eff}$):}
\item \footnotesize{\textit{BVI} filters, $\lambda_{eff}$ =~439.2, 578.6 and 817.6 nm,} 
\item \footnotesize{\textit{JHK} filters, $\lambda_{eff}$ =~1.24, 1.62 and 2.13 $\upmu$m.}
\end{tablenotes}
\end{threeparttable}
\end{table*}

\begin{table*}
\begin{threeparttable}
\caption{TUG optical measurements.}
\label{tab:tug}
\begin{tabular}{cccccccc}
\toprule 
Date & \textit{B} mag  & Date & \textit{V} mag & Date & \textit{R} mag & Date & \textit{I} mag\\
\midrule
397.7037& 14.783 $\pm$ 0.006& 389.7978& 14.252 $\pm$ 0.004& 389.7987& 14.205 $\pm$ 0.005& 389.7994& 13.892 $\pm$ 0.015\\
397.7052& 14.826 $\pm$ 0.006& 389.8000& 14.239 $\pm$ 0.005& 389.8009& 14.225 $\pm$ 0.005& 389.8016& 13.881 $\pm$ 0.015\\
397.7060& 14.838 $\pm$ 0.005& 389.8022& 14.257 $\pm$ 0.004& 389.8031& 14.238 $\pm$ 0.006& 389.8038& 13.864 $\pm$ 0.015\\
397.7088& 14.826 $\pm$ 0.005& 389.8045& 14.250 $\pm$ 0.004& 389.8053& 14.232 $\pm$ 0.006& 389.8060& 13.876 $\pm$ 0.015\\
397.7110& 14.840 $\pm$ 0.004& 389.8067& 14.253 $\pm$ 0.004& 389.8075& 14.237 $\pm$ 0.006& 389.8082& 13.901 $\pm$ 0.016\\
397.7126& 14.840 $\pm$ 0.004& 389.8089& 14.242 $\pm$ 0.004& 389.8097& 14.229 $\pm$ 0.011& 389.8104& 13.859 $\pm$ 0.015\\
398.6950& 14.514 $\pm$ 0.006& 389.8111& 14.242 $\pm$ 0.005& 389.8119& 14.232 $\pm$ 0.007& 389.8126& 13.877 $\pm$ 0.017\\
398.6961& 14.482 $\pm$ 0.006& 389.8133& 14.260 $\pm$ 0.004& 389.8141& 14.211 $\pm$ 0.009& 389.8148& 13.874 $\pm$ 0.016\\
398.6969& 14.501 $\pm$ 0.006& 389.8155& 14.242 $\pm$ 0.004& 389.8164& 14.229 $\pm$ 0.007& 389.8170& 13.886 $\pm$ 0.016\\
$-$& $-$& 389.8177& 14.263 $\pm$ 0.004& 389.8186& 14.220 $\pm$ 0.014& 389.8192& 13.916 $\pm$ 0.021\\
$-$& $-$& 393.7985& 14.457 $\pm$ 0.004& 393.7994& 14.454 $\pm$ 0.007& 393.8000& 14.137 $\pm$ 0.017\\
$-$& $-$& 393.8007& 14.461 $\pm$ 0.005& 393.8016& 14.451 $\pm$ 0.006& 393.8022& 14.148 $\pm$ 0.016\\
$-$& $-$& 393.8029& 14.469 $\pm$ 0.004& 393.8038& 14.465 $\pm$ 0.006& 393.8044& 14.120 $\pm$ 0.015\\
$-$& $-$& 393.8051& 14.463 $\pm$ 0.004& 393.8060& 14.446 $\pm$ 0.005& 393.8066& 14.096 $\pm$ 0.016\\
$-$& $-$& 393.8073& 14.459 $\pm$ 0.004& 393.8082& 14.446 $\pm$ 0.005& 393.8088& 14.120 $\pm$ 0.016\\
$-$& $-$& 393.8095& 14.475 $\pm$ 0.004& 393.8104& 14.468 $\pm$ 0.006& 393.8111& 14.121 $\pm$ 0.016\\
$-$& $-$& 393.8117& 14.476 $\pm$ 0.005& 393.8126& 14.469 $\pm$ 0.007& 393.8133& 14.143 $\pm$ 0.016\\
$-$& $-$& 393.8139& 14.485 $\pm$ 0.004& 393.8148& 14.477 $\pm$ 0.007& 393.8155& 14.130 $\pm$ 0.016\\
$-$& $-$& 393.8161& 14.469 $\pm$ 0.004& 393.8170& 14.481 $\pm$ 0.006& 393.8177& 14.140 $\pm$ 0.016\\
$-$& $-$& 393.8184& 14.466 $\pm$ 0.004& 393.8192& 14.477 $\pm$ 0.006& 393.8199& 14.142 $\pm$ 0.016\\
$-$& $-$& 397.7000& 14.473 $\pm$ 0.004& 397.6944& 14.434 $\pm$ 0.006& 397.6972& 13.910 $\pm$ 0.016\\
$-$& $-$& 397.7011& 14.486 $\pm$ 0.005& 397.6953& 14.440 $\pm$ 0.007& 397.6980& 13.985 $\pm$ 0.016\\
$-$& $-$& 397.7019& 14.492 $\pm$ 0.004& 397.6961& 14.415 $\pm$ 0.007& 397.6987& 14.006 $\pm$ 0.017\\
$-$& $-$& 398.6916& 14.168 $\pm$ 0.004& 397.7174& 14.487 $\pm$ 0.009& 398.6887& 13.661 $\pm$ 0.016\\
$-$& $-$& 398.6927& 14.149 $\pm$ 0.004& 398.6860& 14.128 $\pm$ 0.007& 398.6894& 13.574 $\pm$ 0.016\\
$-$& $-$& 398.6937& 14.171 $\pm$ 0.004& 398.6871& 14.131 $\pm$ 0.005& 398.6901& 13.599 $\pm$ 0.015\\
$-$& $-$& 405.6867& 14.098 $\pm$ 0.005& 398.6878& 14.112 $\pm$ 0.007& 405.6884& 13.532 $\pm$ 0.016\\
$-$& $-$& 405.6891& 14.062 $\pm$ 0.004& 405.6877& 13.949 $\pm$ 0.005& 405.6908& 13.478 $\pm$ 0.015\\
$-$& $-$& 405.6915& 14.061 $\pm$ 0.004& 405.6901& 13.976 $\pm$ 0.004& 405.6932& 13.471 $\pm$ 0.015\\
$-$& $-$& 405.6939& 14.077 $\pm$ 0.004& 405.6925& 13.948 $\pm$ 0.005& 405.6956& 13.494 $\pm$ 0.015\\
$-$& $-$& 405.6963& 14.064 $\pm$ 0.004& 405.6948& 13.952 $\pm$ 0.005& 405.6979& 13.474 $\pm$ 0.015\\
$-$& $-$& 405.6987& 14.063 $\pm$ 0.005& 405.6972& 13.978 $\pm$ 0.005& 405.7003& 13.490 $\pm$ 0.015\\
$-$& $-$& 405.7011& 14.062 $\pm$ 0.004& 405.6996& 13.931 $\pm$ 0.006& 405.7027& 13.473 $\pm$ 0.016\\
$-$& $-$& 405.7035& 14.071 $\pm$ 0.004& 405.7020& 13.926 $\pm$ 0.006& 405.7051& 13.476 $\pm$ 0.016\\
$-$& $-$& 405.7059& 14.081 $\pm$ 0.004& 405.7044& 13.959 $\pm$ 0.006& 405.7075& 13.433 $\pm$ 0.015\\
$-$& $-$& 405.7083& 14.073 $\pm$ 0.004& 405.7068& 13.963 $\pm$ 0.006& 405.7099& 13.489 $\pm$ 0.015\\
$-$& $-$& 407.7967& 14.084 $\pm$ 0.005& 405.7092& 13.960 $\pm$ 0.006& 407.7982& 13.513 $\pm$ 0.015\\
$-$& $-$& 407.7989& 14.084 $\pm$ 0.004& 407.7976& 13.987 $\pm$ 0.007& 407.8004& 13.496 $\pm$ 0.016\\
$-$& $-$& 407.8011& 14.106 $\pm$ 0.004& 407.7998& 13.995 $\pm$ 0.007& 407.8026& 13.522 $\pm$ 0.017\\
$-$& $-$& 407.8033& 14.106 $\pm$ 0.004& 407.8020& 14.002 $\pm$ 0.008& 411.7116& 13.645 $\pm$ 0.017\\
$-$& $-$& 411.7094& 14.217 $\pm$ 0.004& 411.7106& 14.090 $\pm$ 0.008& 411.7146& 13.597 $\pm$ 0.016\\
$-$& $-$& 411.7127& 14.298 $\pm$ 0.005& 411.7138& 14.136 $\pm$ 0.008& 411.7181& 13.719 $\pm$ 0.019\\
$-$& $-$& 411.7154& 14.200 $\pm$ 0.004& 411.7173& 14.089 $\pm$ 0.009& 417.6825& 14.003 $\pm$ 0.019\\
$-$& $-$& 417.6803& 14.490 $\pm$ 0.004& 417.6847& 14.467 $\pm$ 0.015& 417.6857& 13.952 $\pm$ 0.021\\
$-$& $-$& 417.6836& 14.447 $\pm$ 0.004& 417.6880& 14.305 $\pm$ 0.017& 417.6897& 14.057 $\pm$ 0.018\\
$-$& $-$& 417.6868& 14.539 $\pm$ 0.004& 417.6889& 14.452 $\pm$ 0.018& 417.6928& 14.010 $\pm$ 0.016\\
$-$& $-$& 417.6908& 14.415 $\pm$ 0.005& 417.6919& 14.416 $\pm$ 0.009& 420.6790& 14.064 $\pm$ 0.017\\
$-$& $-$& 420.6775& 14.643 $\pm$ 0.004& 420.6783& 14.575 $\pm$ 0.007& 420.6812& 14.091 $\pm$ 0.016\\
$-$& $-$& 420.6797& 14.612 $\pm$ 0.004& 420.6805& 14.514 $\pm$ 0.009& 420.6834& 14.143 $\pm$ 0.018\\
$-$& $-$& 420.6819& 14.549 $\pm$ 0.004& 420.6827& 14.591 $\pm$ 0.008& 420.6856& 14.100 $\pm$ 0.017\\
$-$& $-$& 420.6841& 14.595 $\pm$ 0.004& 420.6850& 14.571 $\pm$ 0.010& $-$& $-$\\
\bottomrule
\end{tabular}
\begin{tablenotes}
\item \footnotesize{Note: $^{*}$Date = MJD - 58000 in days}
\item \footnotesize{Note: Effective wavelength for the R filter, $\lambda_{eff}$= ~634.9 nm.}
\end{tablenotes}
\end{threeparttable}
\end{table*}


\bsp	
\label{lastpage}
\end{document}